\documentclass[namedreferences]{SolarPhysics}
\usepackage[optionalrh]{spr-sola-addons} 
\usepackage{graphicx}        
\usepackage{color}           
\usepackage{url}             

\def\rmit#1{{\it #1}}              

\newcommand\ie{\rmit{i.e.}}
\newcommand\eg{\rmit{e.g.}}

\newcommand\kms{\hbox{km$\;$s$^{-1}$}}
\newcommand\ms{\hbox{m$\;$s$^{-1}$}}

\begin{document}

\begin{article}

\begin{opening}

\title{Non-Linear Numerical Simulations of Magneto-Acoustic Wave
Propagation in Small-Scale Flux Tubes}

\author{E.~\surname{Khomenko}$^{1,2}$\sep
        M.~\surname{Collados}$^{1}$\sep
        T.~\surname{Felipe}$^{1}$
       }
\runningauthor{E. Khomenko, M. Collados and T. Felipe}
\runningtitle{Simulations of MHD Waves in Flux Tubes}

   \institute{$^{1}$ Instituto de Astrof\'{\i}sica de Canarias, 38205, C/ V\'{\i}a
L{\'a}ctea, s/n, Tenerife, Spain; email: \url{khomenko@iac.es, mcv@iac.es, tobias@iac.es} \\
              $^{2}$ Main Astronomical Observatory, NAS, 03680 Kyiv, Zabolotnogo
str. 27, Ukraine \\
             }

\begin{abstract}
We present results of non-linear, 2D, numerical simulations of
magneto-acoustic wave propagation in the photosphere and
chromosphere of small-scale flux tubes with internal structure.
Waves with realistic periods of three to five minutes are studied,
after applying horizontal and vertical oscillatory perturbations
to the equilibrium model. Spurious reflections of shock waves from
the upper boundary are minimized thanks to a special boundary
condition. This has allowed us to increase the duration of the
simulations and to make it long enough to perform a statistical
analysis of oscillations. The simulations show that deep
horizontal motions of the flux tube generate a slow (magnetic)
mode and a surface mode. These modes are efficiently transformed
into a slow (acoustic) mode in the $v_A < c_S$ atmosphere. The
slow (acoustic) mode propagates vertically along the field lines,
forms shocks and remains always within the flux tube. It might
deposit effectively the energy of the driver into the
chromosphere. When the driver oscillates with a high frequency,
above the cut-off, non-linear wave propagation occurs with the
same dominant driver period at all heights. At low frequencies,
below the cut-off, the dominant period of oscillations changes
with height from that of the driver in the photosphere to its
first harmonic (half period) in the chromosphere. Depending on the
period and on the type of the driver, different shock patterns are
observed.
\end{abstract}
\keywords{Photosphere, Chromosphere, Magnetohydrodynamics,
Oscillations, Magnetic fields}
\end{opening}

\section{Introduction}
     \label{sect:Introduction}

Solar magnetic structures show a continuous distribution of fluxes
and sizes. In active regions, the field strength concentrated in
large-scale structures (sunspots and pores) can be as large as
$2-4$ kG. In plage and network areas, the average flux decreases
to several hundreds Gauss. However, individual flux tubes in these
areas can have intrinsic magnetic field strength similar to that
of solar pores, \ie\ $1-2$ kG. The decrease of the average flux is
caused mainly by the decrease of the size of the magnetic
features, \ie\ their filling factor.

The observed photospheric brightness of the magnetic structures is
in close relationship with their flux and size (see for example
\opencite{Berger+etal2007}). Large-scale features like sunspots
and pores appear dark in photospheric observations.
Plage and facular flux tubes appear as bright features (see for
example Figure 7 in \opencite{Berger+etal2007}).

Due to their larger size, sunspots show horizontal gradients of
the magnetic field strength and inclination. Instead, small-scale
flux tubes possess a more homogeneous field with a rather sharp
transition between the magnetized and non-magnetized surroundings.
From the point of view of wave propagation, it is important to
realize that the temperature at a given geometrical height is
different in the different structures. Temperature and magnetic
field define the characteristic wave propagation speeds, \ie\
sound speed, $c_S$, and Alfv\'en speed, $v_A$, as well as the
acoustic cut-off frequency, $\omega_C$, and the height of the
transformation layer where $c_S$ is equal to $v_A$ and different
wave modes can interact.
These parameters can be rather different in large-scale dark
structures and small-scale bright structures.
Thus, it is not surprising  that the observed properties of waves
in sunspots, network, and plage areas are rather different.

\subsection{Sunspots and Pores}

Several decades of studies of sunspot oscillations can be
summarized as follows. Waves observed at different layers of the
atmosphere of sunspots umbrae and penumbrae seem to be the
manifestation of the same phenomenon (see \eg\/
\opencite{Maltby+etal1999}, \citeyear{Maltby+etal2001},
\opencite{Brynildsen+etal2000}, \citeyear{Brynildsen+etal2002},
\opencite{Christopoulou+etal2000},
\citeyear{Christopoulou+etal2001}, \opencite{Rouppe+etal2003},
\opencite{Tziotziou+etal2006}).
In the photosphere, five-minute oscillations are observed in
sunspot umbrae and penumbrae and in solar pores
\cite{Bogdan+Judge2006}. Wave propagation is linear with
amplitudes of several hundreds of \ms\ and is parallel to the
magnetic field lines (see \eg\/ \opencite{Gurman+Leibacher1984},
\opencite{Lites1984}, \opencite{Collados2001},
\opencite{Centeno+etal2006a}, \opencite{Bloomfield+etal2007b}).
Magnetic field oscillations of a few Gauss are detected in the
deeper layers and attempts have been made to interpret these
oscillations in terms of fast and slow MHD waves (\eg\/
\opencite{Ruedi+Solanki+Stenflo+Tarbell+Scherrer1998},
\opencite{Lites+Thomas+Bogdan+Cally1998},
\opencite{BellotRubio+Collados+RuisCobo+RodriguezHidalgo2000},
\opencite{Khomenko+Collados+BellotRubio2003}).

In the chromosphere, the dominant period of oscillations decreases
and shock waves are observed with a three-minute periodicity in
sunspot umbrae and in pores (\opencite{Lites1984},
\citeyear{Lites1986}, \citeyear{Lites1988},
\opencite{Socas-Navarro+etal2000}, \opencite{Centeno+etal2006a}
\opencite{Tziotziou+etal2007}). The amplitudes of the observed
shocks are in the range of $5-15$ \kms, depending on the size of
the structure. In pores, shocks are weaker
\cite{Centeno+etal2006b}. Umbral flashes also show a three-minute
periodicity. In contrast, in the penumbra, the shock behavior of
waves with five-minute periodicity is detected for running
penumbral waves even at chromospheric heights
\cite{Tziotziou+etal2006}.
Waves observed in the transition region and the corona conserve
these properties (\opencite{DeMoortel+etal2002},
\opencite{Marsh+Walsh2005}, \citeyear{Marsh+Walsh2006},
\opencite{Maltby+etal1999}, \citeyear{Maltby+etal2001},
\opencite{Brynildsen+etal2000}, \citeyear{Brynildsen+etal2002},
\opencite{Christopoulou+etal2000},
\citeyear{Christopoulou+etal2001}).
Wave propagation from the photospheric pulse into the chromosphere
and higher layers is along the magnetic field lines for both three
and five minutes perturbations.

\subsection{Plages, Facular regions, and Network}


Bright magnetic field structures like plages, facular regions, and
network manifest five-minute linear oscillations in the
photosphere and the oscillations maintain the five minute
periodicity in the chromosphere and larger heights
(\opencite{Lites+Rutten+Kalkofen1993}, \opencite{Krijer+etal2001},
\opencite{DePontieu+etal2003}, \opencite{Centeno+etal2006b},
\opencite{Bloomfield+etal2006}, \opencite{Veccio+etal2007}).
Shock occurrence is much lower in these quiet Sun regions with
enhanced flux (\opencite{Krijer+etal2001},
\opencite{Centeno+etal2006b}).
The velocity oscillation amplitudes in the chromosphere do not
reach values as large as in sunspots and stay within $2-4$ \kms.
Waves at photospheric and chromospheric layers seem to be
correlated and vertical propagation along the vertical magnetic
fields seem to dominate (\opencite{Krijer+etal2001},
\opencite{Centeno+etal2006b}, \opencite{Veccio+etal2007}).
Curiously, there are areas with enhanced three-minute power
surrounding the strong magnetic field concentrations, possibly
related to flux tube canopies (aureoles, see
\opencite{Braun+Linsday1999}, \opencite{Thomas+Stanchfield2000},
\opencite{Krijer+etal2001}, \opencite{Judge+etal2001}).
In the higher coronal layers, loops connecting sunspots oscillate
with three minutes and the loops connecting network and plage
regions oscillate with five minute periods
\cite{DeMoortel+etal2002}.

Thus, the wave properties are distinctly different in large-scale
and small-scale magnetic features. The difference in oscillation
properties is present through the whole solar atmosphere and
should be a consequence of the thermal and magnetic properties of
these structures.

\subsection{Theoretical Models}

A number of theoretical mechanisms have been proposed to explain
the oscillation spectra at different heights in magnetic
structures and to interpret the observed oscillations in terms of
MHD waves.
\inlinecite{Zhugzhda+Locans1981},
\inlinecite{Gurman+Leibacher1984}, and \inlinecite{Zhugzhda2007}
argue that the observed spectrum of sunspot umbral oscillations in
the chromosphere is due to the temperature gradients of the
atmosphere acting as an interference filter for linear
three-minute period acoustic waves.
On the other hand, it has been demonstrated by
\inlinecite{Fleck+Schmitz1991} that the change of the period with
height from five to the three minutes is a basic phenomena
occurring even in an isothermal atmosphere for linear waves due to
the resonant excitation at the atmospheric cut-off frequency. In
the case of the solar atmosphere, the temperature minimum gives
rise to the three-minute cut-off period.
Later, the same authors found that, due to non-linear shock
interaction and overtaking, the frequency of acoustic oscillations
also changes with height (\opencite{Fleck+Schmitz1993}). The
response of the solar atmosphere to an adiabatic shock wave leads
to a natural appearance of the three-minute peak in the
oscillation power spectra, in the case the underlying photosphere
has a five-minute periodicity.
A third mechanism was shown to explain the three-minute
periodicity of calcium bright grains in the quiet Sun: with a
spectrum of periods in the photosphere with a peak at five
minutes, this distribution changes to one with a maximum at the
acoustic cut-off because at longer periods the energy falls off
exponentially with height (\opencite{Carlsson+Stein1997}).
These mechanisms may explain the shift with height of the period
of oscillations from five to three minutes observed in sunspots.

The question remains: why is there not such a shift in small-scale
structures of plages and network regions?
\inlinecite{DePontieu+etal2004} argue that the inclination of the
magnetic field may play an important role. If (acoustic or slow
MHD) waves have a preferred direction of propagation defined by
the magnetic field, the effective cut-off frequency is lowered by
the cosine of the inclination angle with respect to the vertical.
This allows evanescent waves to propagate.
It should be recalled, however, that mainly vertical propagation
is observed in the photosphere and chromosphere in plage regions
(see \eg\/ \opencite{Krijer+etal2001},
\opencite{Centeno+etal2006b}), thus it is unclear whether the
mechanism suggested by \inlinecite{DePontieu+etal2004} is at work.
Alternatively, the decrease of the effective acoustic cut-off
frequency can be produced by taking into account the radiative
losses of acoustic oscillations \cite{Roberts1983,
Centeno+etal2006b}. If the radiative relaxation time is
effectively small (as expected in small-scale magnetic structures,
transparent to radiation), the cut-off frequency can be reduced
and evanescent waves can propagate.

All works cited above apply the theory of acoustic waves in
idealized atmospheres in order to explain the observed oscillation
properties in magnetic structures.
In magnetized atmospheres, different wave types can exist and
different modes can be observed depending on the magnetic field
configuration and the height of the transformation layer
($c_S=v_A$) relative to the height of the formation of the
spectral lines used in observations.
\inlinecite{Bogdan+etal2003}, \inlinecite{Rosenthal+etal2002},
\inlinecite{Hasan+Ulmschneider2004}, Hasan {\it et al.} (2003,
2005), \nocite{Hasan+etal2003} \nocite{Hasan+etal2005} addressed
this question in their simulations of magneto-acoustic waves in
non-trivial magnetic configurations.
They pointed out the importance of defining the mode type and the
mode conversion height when interpreting the observations. Due to
mode mixing at the transformation layer, no correlation may be
observed between perturbations below and above the $c_S=v_A$
height.

All these arguments point of towards the need to have realistic,
self-consistent, and systematic models of the wave spectrum in
magnetic structures. It is necessary to include non-trivial
magnetic field configurations in order to allow the existence of
different mode types and mode conversion. Non-linearities should
be taken into account for waves propagating into the chromosphere.
The thermal structure should be realistic in order to make
possible the reflection of waves with frequencies below the
cut-off. Radiative damping also plays an important role, in
particular to estimate the amount of energy that may deposited in
the higher atmosphere due to waves.
In the present work we address several of the above questions. We
perform non-linear numerical modeling of magneto-acoustic waves in
small-scale magnetic flux tubes with internal structure. We
consider waves with realistic periods of three and five minutes,
which has never been done before for non-trivial magnetic
configurations.
We study the mode transformation and oscillation spectra at
different heights from 0 to 2000 km as a function of the type of
the driver and its period.
At present, radiative damping of oscillations is not taken into
account. This question will be addressed in a separate paper.

\begin{figure}
\centerline{\includegraphics[width=1.0\textwidth]{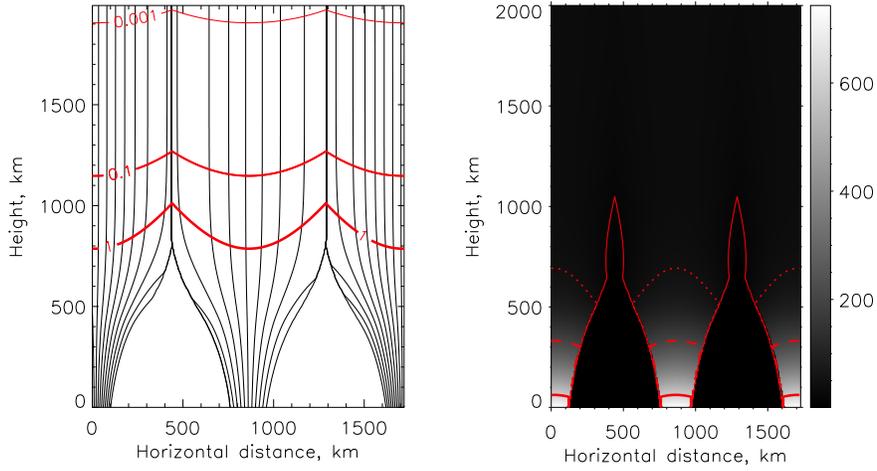}}
              \caption{Second-order flux tube magnetostatic model, calculated
              after Pneuman, Solanki, and Stenflo (1986). The tube radius
              is equal to 100 km in the photosphere. Left
              panel: magnetic field lines and contours of plasma
              $\beta$ (defined as $c_S^2/v_A^2$), with values indicated on each
              contour. Right panel: distribution of the magnetic field strength.}
\label{fig:tube}
\end{figure}

\section{Numerical Model}

We solve the basic equations of the ideal MHD, written in
conservative form:
\begin{equation}
 \frac{\partial \rho }{\partial \it{t}} +  \vec{\nabla} \cdot
(\rho \vec{V})=  0 \,, \label{eq:den}
\end{equation}
\begin{equation}
\frac{\partial (\rho \vec{V}) }{\partial \it{t}} + \vec{\nabla}
\cdot [ \rho \vec{V}\vec{V} + (P + \frac{\vec{B}^2}{8\pi}) {\bf I}
- \frac{\vec{B}\vec{B}}{4\pi}]=\rho\vec{g} \,, \label{eq:mom}
\end{equation}
\begin{equation}
\frac{\partial E}{\partial \it{t}} + \vec{\nabla}\cdot[(E + P +
\frac{\vec{B}^2}{8\pi})\vec{V} - \vec{B}(\frac{\vec{B} \cdot
\vec{V}}{4\pi})] = \rho\vec{V}\cdot\vec{g} + \rho Q  \,,
\label{eq:ene}
\end{equation}
\begin{equation}
 \label{eq:ind}
\frac{\partial \vec{B}}{\partial \it{t}} =
\vec{\nabla}\times(\vec{V} \times \vec{B})  \,,
 \end{equation}
where ${\bf I}$ is the diagonal identity tensor and $E$ is the
total energy:
\begin{equation}
E=\frac{1}{2}\rho V^2 + \frac{P}{\gamma-1} + \frac{B^2}{8\pi} \,.
 \end{equation}
All other symbols have their usual meaning.
The numerical code solving these equations was used previously in
\inlinecite{Khomenko+Collados2006} and is described in details in
that paper. For the present work, the code has been improved to
solve the fully non-linear equations. One of the main differences
between our code and any other standard MHD code is that we solve
equations for perturbations (including all non-linear terms), the
magnetostatic equilibrium being analytically subtracted from the
above system of equations.

The term describing energy losses, ($Q$), is set to zero in the
present study.
A Perfect Matching Layer (PML) boundary condition
\cite{Berenger1994} is applied to the sides and top boundaries of
the simulation domain. It is found to perform rather well in our
simulations. Tests have shown that wave reflection from the upper
boundary can be reduced to a few percent even for shock waves with
Mach number of two or three. This has allowed us to perform longer
temporal series of simulations, not stopping them at the time when
the fastest wave reaches the upper boundary, and to analyze these
simulations in a statistical way, similar to observations.
The boundary conditions and other numerical details are described
elsewhere \cite{Khomenko+Collados2006}.
The simulations analyzed below are done in two dimensions.

At the bottom (photospheric) boundary, we specify either a
vertical or horizontal velocity as a function of time and
horizontal coordinate ($x$):
\begin{eqnarray}
V_{x,z}(x,\it{t})=V_0 \sin(\omega \it{t})\times
\exp(-(x-x_0)^2/2\sigma^2) \,. \label{eq:v0}
\end{eqnarray}
The perturbation is localized in the $x$-axis and has a Gaussian
shape in this direction.
The horizontal size of the pulse $\sigma=160$ km, initial
amplitude at the photosphere $V_0=200$ \ms\ and $x_0$ corresponds
to the tube axis.

\section{Magnetostatic Model}

In order to construct an initial magnetostatic structure, we have
followed the method described in \inlinecite{Pneuman+etal1986}.
The equations for magnetostatic equilibrium have been written in
cylindrical coordinates. All the variables were expanded in a
power series in the horizontal radial coordinate $r$. The series
have been truncated after the second order in $r$. After some
manipulations with the equations, one single equation can be
obtained for the vertical stratification of the magnetic field
strength at the flux tube axis (for details, see
\opencite{Pneuman+etal1986}). This equation has been solved
numerically using an iterative interpolation scheme from
\inlinecite{Korn}. Once the magnetic field at the axis is
obtained, all of the other variables can be calculated from
analytical expressions.
The free parameters of the model are: magnetic field strength and
flux tube radius at the base of photosphere and filling factor.
The latter parameter allows us to produce a smooth merging of flux
tubes at some chromospheric height. By varying these parameters,
we have obtained a solution with desired properties concerning the
height of the transformation layer $c_S=v_A$, inclination of field
lines, and size of the tube. The temperatures inside and outside
the flux tube were kept constant at all heights. The initial
stratification of the thermodynamic parameters were taken from the
VAL-C atmospheric model \cite{Vernazza+Avrett+Loeser1981}.
Table 1 and Figure~\ref{fig:tube} summarize the magnetic and
thermal properties of our magnetostatic solution.

\begin{table}
\caption{ Characteristic parameters of the model flux tube taken
at the flux tube axis. } \label{T-simple}
\begin{tabular}{lcc}     
  \hline                   
Parameter & z=0 km & z=2000 km  \\
  \hline
Temperature (K) & 6420 & 7175   \\
Magnetic field (G) & 743 & 37   \\
Sound speed (\kms) & 8 & 10 \\
Alfven speed (\kms) & 4 & 394 \\
$c_S^2/v_A^2$ & 4 & 7$\times 10^{-4}$ \\
Pressure scale height (km) &    151    & 228       \\
Cut-off period (s) &    230    &  280      \\
  \hline
\end{tabular}
\end{table}

Since our model flux tube is a second-order approximation to a
thin flux tube, it has horizontal variations of magnetic field
strength, pressure, {\it etc.} This approximation also gives rise
to horizontal and vertical variations of the plasma $\beta$ (note
that through the present paper we define $\beta$ as
$c_S^2/v_A^2$). The transition between the outside and inside
atmospheres is rather sharp. Due to pressure balance, the gas
pressure is smaller inside than outside the flux tube. The
magnetic field lines fan out until a certain height, where nearby
flux tubes merge together. After that height, the magnetic field
becomes vertical and almost homogeneous in the horizontal
direction.

The magnetostatic model shown in Figure~\ref{fig:tube} was
perturbed by either a vertical or a horizontal driver, as follows
from Equation (\ref{eq:v0}). The results of the simulations for
different driving periods are described in the remainder of the
paper.

\section{``Mode Conversion'' Nomenclature}

There is a varying usage of the term ``mode conversion'' both in
astrophysics and in plasma physics literature.

On the one side, the term ``mode conversion'' is used when the
mode changes from slow to fast or the other way around (see, \eg\/
\inlinecite{Cally2006}). The transformation of a slow (magnetic)
mode into slow (acoustic) mode is not called ``mode conversion'',
but instead ``the changing of the character of slow mode'' (same
for the fast mode).

On the other side, in a later paper by \inlinecite{Cally2007}, the
terms ``conversion'' and ``transmission'' are used. The term
``conversion'' is used to describe the change from acoustic to
magnetic nature of the wave or the other way around, \ie\,
slow--slow or fast--fast. The term  ``transmission'' is used to
describe the change from slow to fast or from fast to slow. This
terminology used in parts of the plasma physics literature as
well.

Since there seems to be no standard to describe what a mode
conversion is, we find it necessary to define the nomenclature we
utilize below in the paper. We use the term ``mode conversion'' in
a more general sense. When the propagation speed of wave changes
from fast to slow, we call it ``mode conversion from fast mode to
slow mode'' (the same for the opposite). When the physics of the
restoring force changes from magnetic to acoustic, we call it
``mode conversion from slow (magnetic) mode to slow (acoustic)
mode'' (the same for the fast mode).

\section{Horizontal Driving at 50 Seconds}

\begin{figure}
\centerline{\includegraphics[width=1.0\textwidth]{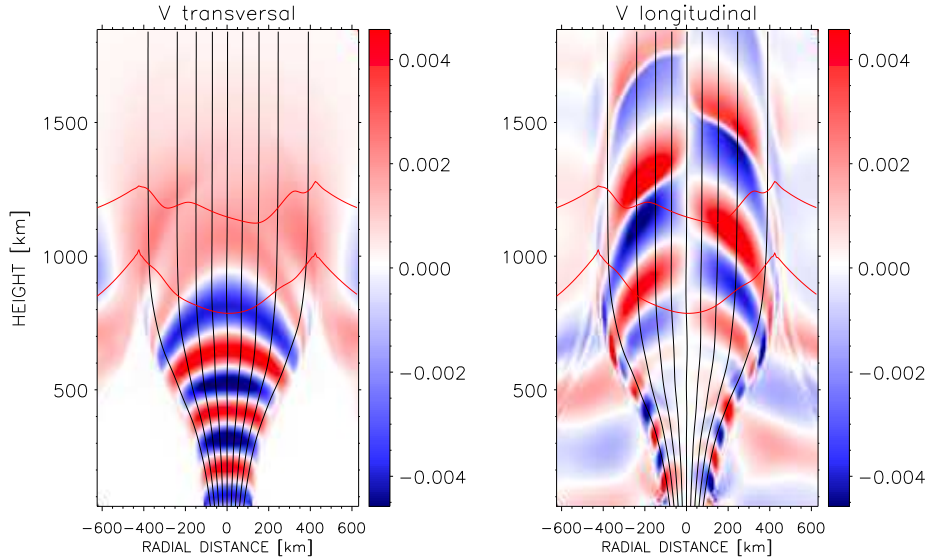}}
              \caption{Snapshot of the simulations excited with a horizontal
driver at 50 seconds. The panels show the transverse
(perpendicular to the magnetic field) and longitudinal (parallel
to the magnetic field) velocity components at an elapsed time of
340 seconds after the start of the simulations. The transverse
component is normalized to $\sqrt{v_A \rho_0}$ and the
longitudinal component is normalized to $\sqrt{c_S \rho_0}$ in
order to highlight the variations of the energy flux as a function
of height. The black lines represent the magnetic field lines. The
two red lines are contours of constant $c_S^2/v_A^2$, the thick
line corresponding to $v_A=c_S$ and the thin line to
$c_S^2/v_A^2=0.1$. Note the change of the position of both curves,
compared to Figure 1, produced by wave motions.}
\label{fig:snap50}
\end{figure}

\begin{figure}
\centerline{\includegraphics[height=0.95\textheight]{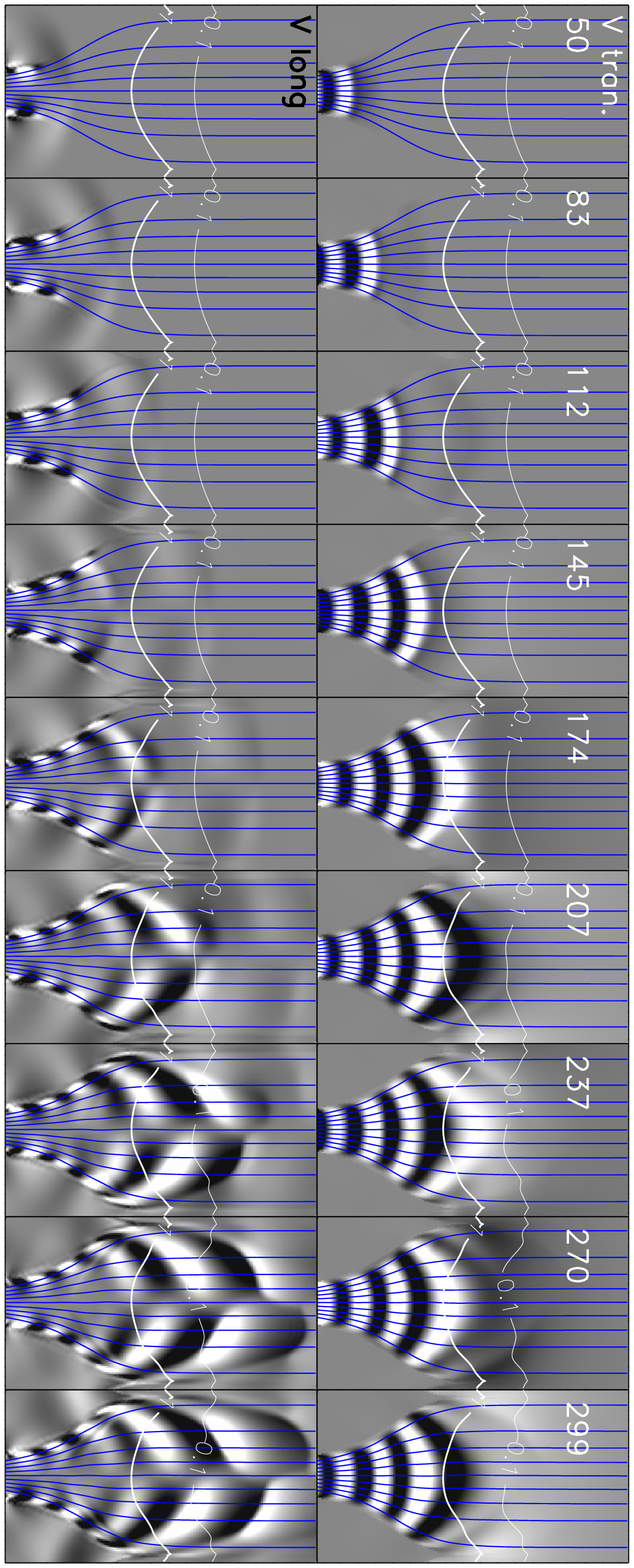}}
              \caption{Time series of snapshots of the transverse (right)
              and longitudinal (left) velocities in the simulation  with
              a horizontal driver at 50 seconds, showing the early stages of
              the evolution. Numbers give
              the elapsed time in seconds since the start of the simulation.
              The transverse component is normalized to $\sqrt{v_A \rho_0}$
              and the longitudinal component is normalized to $\sqrt{c_S \rho_0}$.
              Each snapshot includes 900
              km in horizontal and 2000 km in vertical directions. }
\label{fig:ft50}
\end{figure}

Simulations of short-period (\ie\ short-wavelength) waves have the
advantage that they make possible to understand easily what types
of wave modes are generated and what transformations take place.

Figure~\ref{fig:snap50} gives a snapshot of the transverse and
longitudinal velocity components at an elapsed time of 340 seconds
after the start of the simulations, and Figure~\ref{fig:ft50}
gives the temporal evolution of this simulation during the early
stages. At each point, the transverse component is normalized to
$\sqrt{v_A \rho_0}$, and the longitudinal component is normalized
to $\sqrt{c_S \rho_0}$. This normalization allows us to compare
the energy fluxes in both components for the case of the linear,
short-wavelength perturbations.

The periodic horizontal driver located at the lower boundary in
the high-$\beta$ regime generates primarily a slow magnetic mode,
detected in the transverse velocity perturbation inside the flux
tube at lower heights.
Due to the Gaussian shape of the driver, the fast mode is also
generated. It can be seen at the early snapshots in
Figure~\ref{fig:ft50} (up to 240 seconds after the start of the
simulation) as a weak perturbation in longitudinal velocity
propagating rapidly up and having a longer wavelength. It can be
seen that the fast wave is rather weak and does not play a major
role at the stationary stage of the simulation. Its contribution
to the total energy flux at the upper atmosphere is only minor.
Note, also, that the fast wave is present outside the flux tube in
the field-free atmosphere. This is due to the fact that the width
of the driver ($\sigma$) is larger than the radius of the flux
tube [Equation(\ref{eq:v0})].

The slow mode generated at the bottom propagates vertically upward
changing the curvature of the wave front due to the horizontal
gradients in the Alfv\'en speed inside the flux tube.
Note that there is no corresponding perturbation in the
longitudinal velocity inside the tube. This confirms that the slow
mode is purely transverse.
In addition to this slow magnetic mode, horizontal motions at the
lower boundary also generate a surface mode. It can be seen very
well in the snapshot of the longitudinal velocity. This mode is
localized on the magnetic--non-magnetic interfaces and follows the
shape of the tube. The perturbation associated with the surface
mode decays exponentially away from the tube boundary inwards and
outwards. The surface mode is longitudinal and can be
distinguished in the simulations until the height where flux tubes
merge and the magnetic field and other parameters become
homogeneous.

In the vicinity of the $c_S=v_A$ layer, both the slow magnetic
mode and the surface mode transfer most of their energy to the
slow acoustic mode. This transformation can be appreciated from
the time evolution of the simulation shown in
Figure~\ref{fig:ft50}, where it is clear that significant
perturbation in the longitudinal velocity in the upper layers
inside the flux tube appears only after the slow mode reaches the
transformation layer. Some of the energy of the slow magnetic mode
also goes into the fast magnetic mode at the $v_A > c_S$ region.
As shown by the sequence of snapshots, this fast mode is refracted
back to the photosphere. However, its energy is small compared to
that of the slow acoustic mode in the $v_A
> c_S$ layers. The ratio of the transverse and longitudinal
velocities at these heights is close to 0.1. This situation should
be contrasted with the simulations of waves in sunspots
\cite{Khomenko+Collados2006}, where a significant part of the
energy of the driver was returned back to the photosphere by the
fast magneto-acoustic mode.
The slow acoustic mode in the $v_A > c_S$ region is longitudinal
and can be observed at heights above \hbox{1000 km} in the
longitudinal velocity snapshot. It propagates along the magnetic
field lines straight up to the chromosphere and carries most of
the energy of the driver to the upper heights.  Motions on
opposite sides of the flux tube are in anti-phase, as a natural
consequence of the horizontal motions of the tube at the
photospheric boundary.

\begin{figure}
\centerline{\includegraphics[width=1.0\textwidth]{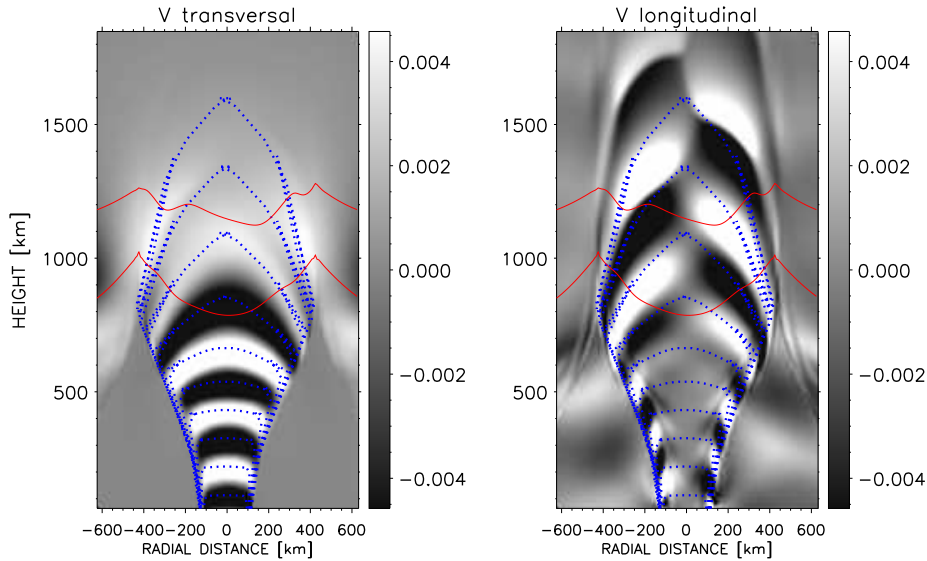}}
              \caption{WKB solution for the simulations with a
 horizontal driver at 50 seconds. The format of the figure is
 the same as Figure~2. Blue contours are the positions of constant
 phase of the WKB slow mode solution at the different equidistant time
 moments. } \label{fig:wkb}
\end{figure}

The finding that the slow magnetic mode is effectively transformed
into the slow acoustic mode is different from the conclusions of,
\eg, \inlinecite{Bogdan+etal2003} who obtain from their simulation
that a significant part of the slow-mode energy goes into the fast
mode, thus preserving the magnetic nature of the wave at all
heights.
In order to better understand our simulations, we present in
Figure~\ref{fig:wkb} the WKB solution for the slow mode. The WKB
approach is linear and assumes that the wavelength of the
perturbation is much smaller than the characteristic scale of the
variations of the background atmospheric parameters, and thus can
be applied in the case of the short-wavelength simulations with
the 50-seconds driver. The equations describing the WKB approach
in this particular case can be found in
\inlinecite{Khomenko+Collados2006}. The blue contours in
Figure~\ref{fig:wkb} describe the position of the wave front at
different time moments separated by equal intervals. The initial
wavefront at the base of the atmosphere is assumed planar, \ie\
the horizontal wave vector $k_x$ is equal to zero.
The WKB solution describes rather well the position of the
wavefront at every time moment, \ie\ the waves propagate at the
slow-mode speed.
As can be appreciated from the figure, as the wave propagates
upwards, the horizontal gradients of its phase speed produce the
deformation of the wave front, so that its central part propagates
faster. It means that at greater heights the wave vector
($\vec{k}$) is no longer directed along the vertical, but forms a
significant angle with respect to the magnetic field vector
$\vec{B_0}$. According to \inlinecite{Cally2006}, if the angle
between $\vec{k}$ and $\vec{B_0}$ is close to zero, the most
effective transformation is from the slow to the fast mode.
However, as this angle increases, the slow to slow mode
transformation becomes more important. Since in our simulations
the slow mode reaches the transformation layer with $\vec{k}$
significantly inclined with respect to $\vec{B_0}$, it explains
the effectiveness of the slow to slow mode transformation. This
does not contradict the fact that the wave energy may propagate
longitudinally, which is a result of the strong anisotropy of the
medium.

The slow acoustic mode develops shocks with amplitudes of about 5
\kms\ that propagate straight up into the chromosphere.
Due to the clear non-linear behavior of the acoustic mode at the
greater heights, secondary harmonics are generated in addition to
the frequency of the original perturbation of the driver. However,
the maximum power at all heights is kept at the driving frequency.

\begin{figure}
\centerline{\includegraphics[width=0.95\textwidth]{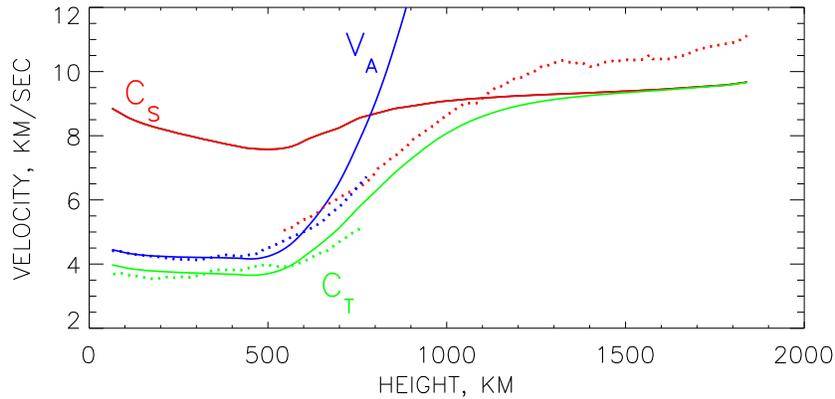}}
              \caption{Height dependence of phase speeds of the slow
              mode (blue dotted line), surface mode (green dotted line)
              and acoustic mode (red dotted line) measured
              from the simulations compared
              to the characteristic propagation speeds calculated
              in the magnetostatic model (solid lines).  The values of
              the Alfv\'en and sound speed ($v_A$ and $c_S$)
              are taken at the flux tube axis. The value of
              the tube speed ($c_T$) is computed along the trajectory of the
              surface mode.}
\label{fig:speeds}
\end{figure}

Figure~\ref{fig:speeds} gives the characteristic speeds of the
waves in the simulation. Using the time series of the simulations,
we have calculated the phase speed of the slow magnetic mode (at
heights from 0 to about 800 km, where this mode exists), the slow
acoustic mode (at heights from 500 km upward) and the surface
mode. In the latter case, the phase speed along the curved
trajectory of the surface mode was calculated. The retrieved phase
speeds are displayed in Figure~\ref{fig:speeds} as a function of
height. The stratifications of $v_A$, $c_S$ and tube speed
($c_T^2=c_S^2 v_A^2 / (c_S^2 + v_A^2)$) are also presented.

The comparison of the dotted and solid line curves in
Figure~\ref{fig:speeds} leads to coherent results. The slow
magnetic mode propagates with the Alfv\'en speed ($v_A$). The
surface mode propagates with the tube phase speed ($c_T$) up to
400 km.
Despite the fact that the tube and the Alfv\'en velocities are
rather similar in the lower photosphere, the phase speed of the
surface wave is distinct from that of the slow magnetic wave.  The
former propagates with its own characteristic speed, as suggested
from the analytical theory (see, \eg\ \opencite{Roberts1981}).
At heights about 700 km (where $v_A \approx c_S$) the speeds of
all modes become close to one another and the energy can be easily
transferred between the different wave types. It can be seen that
the slow magnetic mode and the surface mode give their energy to
the slow acoustic mode.  As the slow acoustic mode propagates
upwards, shock formation occurs and the propagation becomes
supersonic. The phase speed of the shocks increases with height in
agreement with the increase of their amplitudes.

The following items summarize the most important features of the
simulations with the horizontal driver at 50 seconds:
\begin{itemize}
\item Slow to slow mode transformation at the
$v_A=c_S$ height.
\item Left-right acoustic shock wave pattern in anti-phase.
\item Surface mode excitation.
\item The period of the waves is maintained at all
heights.
\end{itemize}

The simulation with this 50 seconds driver follows the spirit of
the calculations by \inlinecite{Rosenthal+etal2002},
\inlinecite{Bogdan+etal2003} and \inlinecite{Hasan+etal2005},
where small scale flux tubes were perturbed by horizontal
high-frequency motions at the lower boundary. Similarly to these
works, we also find that such motions generate acoustic shocks at
chromospheric heights. The differences in the wave behavior
between these works and ours are mainly due to the difference in
the initial magnetostatic situation. Once more, this suggests that
the background magnetostatic state should be taken into account
while interpreting observations, since different wave types can be
generated.

\section{Horizontal Driving at 180 Seconds}

\begin{figure}
\centerline{\includegraphics[height=0.95\textheight]{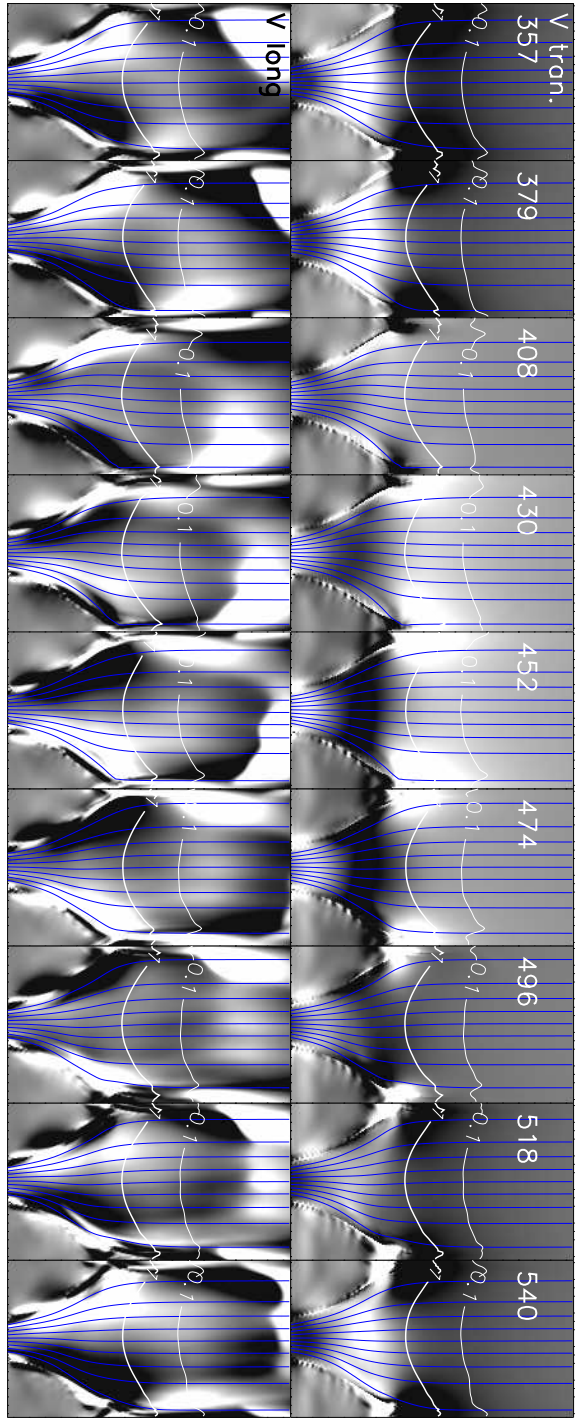}}
              \caption{Time series of snapshots of the transverse (right)
              and longitudinal (left) velocities in the simulation  with
              a horizontal driver at 180 seconds. Numbers give
              the elapsed time in seconds since the start of the simulation.
              After removing the average velocity increase with height produced
              by the density fall-off, all images have the same scale. Each snapshot includes 900
              km in horizontal and 2000 km in vertical direction. }
\label{fig:ft180}
\end{figure}

\begin{figure}
\centerline{\includegraphics[width=0.8\textwidth]{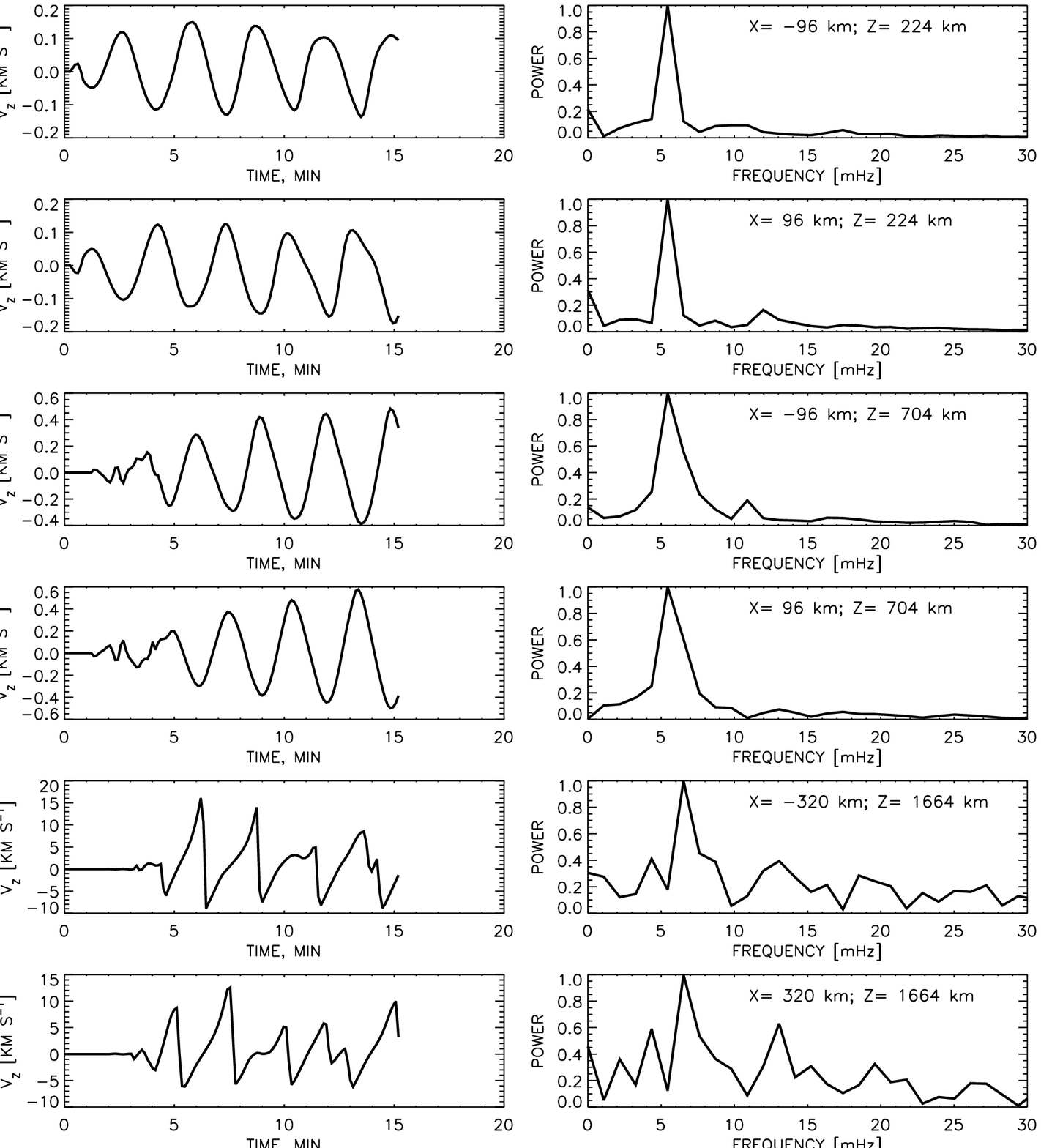}}
              \caption{Examples of the temporal evolution of the longitudinal velocity
              (left) and their corresponding power spectra (right) at
              different heights and positions inside the flux tube for the simulation
              with horizontal driving at 180 seconds. The coordinates
              are indicated on the right panels.}
\label{fig:vz180}
\end{figure}

Here, we show the results of the simulations using a 180-seconds
driver, still above the cut-off frequency. For this reason, it is
not surprising that we get a behavior similar to the previous
case.
The wave pattern that appears in the simulations is basically the
same as in the case of the 50 seconds driver, but the wavelength
of the perturbations is larger.
The driver mainly excites in the lower layers a transverse slow
magnetic mode inside the flux tube, together with a surface mode
at the interface with the non-magnetic surroundings. As before,
the surface mode is mainly longitudinal. Figure~\ref{fig:ft180}
shows some snapshots of the transverse and longitudinal velocities
during one period of the wave after the simulations reached the
stationary state. In contrast to the case of the short-wavelength
50 seconds driver simulation, we find the normalization of the
velocity components to  $\sqrt{c_S \rho_0}$ or $\sqrt{v_A \rho_0}$
to be not appropriate for the 180 and 300 seconds driver
simulations since the group velocity of waves follows a more
complicated behaviour than either $c_S$ or $v_A$ due to the larger
wavelength and non-linear effects.
Instead we normalize both velocity components with an exponential
function ($\rho_0^{1/4}$) that allows a better visualization of
the results.

Around the $c_s = v_A$ layer, mode conversion takes places and
most of the energy goes again to the slow acoustic mode. The
antisymmetric pattern with respect to the flux tube axis is still
present in longitudinal velocity above this layer, although the
antisymmetry is not as perfect as in the 50-seconds driver case.
Shocks clearly develop during the evolution. Very clear examples
are apparent at time 379 in the left part of the flux tube, and at
time 452 in the right part. These non-linear effects are more
pronounced in the higher layers.
The shock formation occurs above a height of about 1000 km.
The temporal evolution of the transverse velocity shows the
refraction of the fast magnetic mode at the higher layers where
$v_A > c_S$.
The horizontal motions of the flux tube field lines in the
photosphere and chromosphere due to this mode can also be seen
during the time evolution.

In Figure~\ref{fig:vz180}, the temporal evolution of the
longitudinal velocity at different points in the flux tube,
corresponding to different heights and horizontal distances from
the axis, is shown.
The first two upper panels on the left correspond to two points
located near the boundary of the flux tube, one on the right and
the other on the left of the axis, 200 km above the driver. The
almost sinusoidal oscillation pattern is clear, since
non-linearities have not had time to develop.
When comparing the position in the horizontal axis of maxima and
minima for the velocity time series from both points, it is
apparent that both points oscillate in anti-phase. The two middle
panels on the left correspond to two points located at the same
distance from the axis than the previous ones, but at a larger
height (704 km).
This is the height where $v_A \approx c_S$ and the wave
transformation starts.
The amplitude of the oscillation has increased by almost a factor
of four. It can be seen how the wave takes some time to reach this
layer.
The perturbation reaches this layer about four minutes after the
driver starts. Up to there, non-linearities are still negligible,
and the sinusoidal aspect of the wave is still kept, as well as
this antisymmetry of the velocity pattern.
At higher layers (1664 km in the plots) velocity discontinuities
appear, indicating the development of shocks, with amplitudes of
several \kms\ (two bottom panels of Figure~\ref{fig:vz180}). The
corresponding power spectra of this velocity temporal series are
plotted in the right-hand part, and they all demonstrate that the
frequency of the driver (\hbox {5.56 Hz}) remains unaltered at all
heights. The presence of several harmonics in higher layers is
indicative of the non-linearities giving rise to the saw-tooth
velocity profiles, rather than sinusoidal.
Note that the shocks are in antiphase at the both sides of the
tube.

Probably, the most important difference with respect to the
previous case with a driver of 50 seconds, lies in the fact that
the slow-to-slow  mode conversion is not as efficient now (see
\opencite{Cally2006}), and a larger part of the energy of the slow
mode excited by the driver is transferred to the fast (transverse,
and hence, mainly magnetic) mode. The atmosphere above the $v_A =
c_S$ oscillates uniformly to left and right with small phase
differences. The horizontal motions are in phase in the whole
atmosphere due to the large oscillation wavelength.

Summarizing,  the most important features of the simulations with
the horizontal driver at 180 seconds are:

\begin{itemize}
\item  Partial slow to slow (acoustic) transformation at $v_A=c_S$ height.
\item  Horizontal flux tube motion at 2000 km due to the fast (magnetic)
wave.
\item  Left-right shock wave pattern in anti-phase.
\item  Surface mode excitation.
\item  Period of the wave maintained with height.
\end{itemize}

\section{Horizontal Driving at 300 Seconds}

\begin{figure}
\centerline{\includegraphics[height=0.95\textheight]{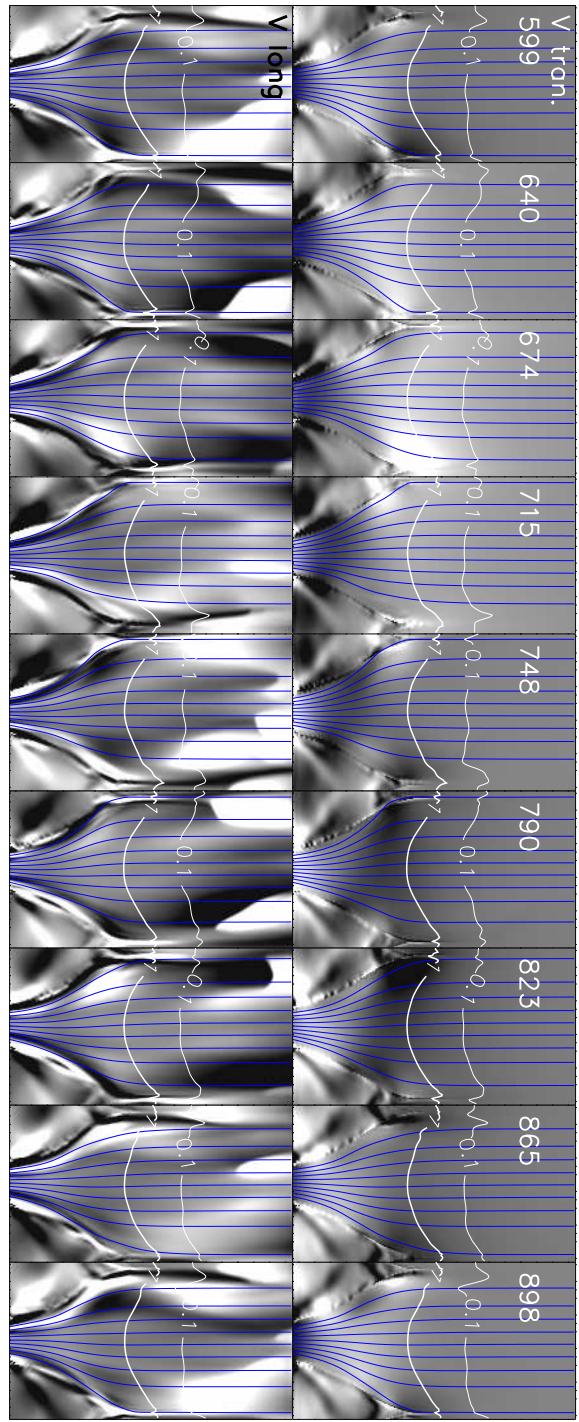}}
              \caption{Time series of snapshots of the transverse (right)
              and longitudinal (left) velocities in the simulation  with
              a horizontal driver at 300 seconds. Numbers give
              the time elapsed in seconds since the start of the simulation.
              After removing the average velocity increase with height produced
              by the density fall-off, all images have the same scale. Each snapshot includes 900
              km in horizontal and 2000 km in vertical direction. }
\label{fig:ft300h}
\end{figure}

\begin{figure}
\centerline{\includegraphics[width=0.8\textwidth]{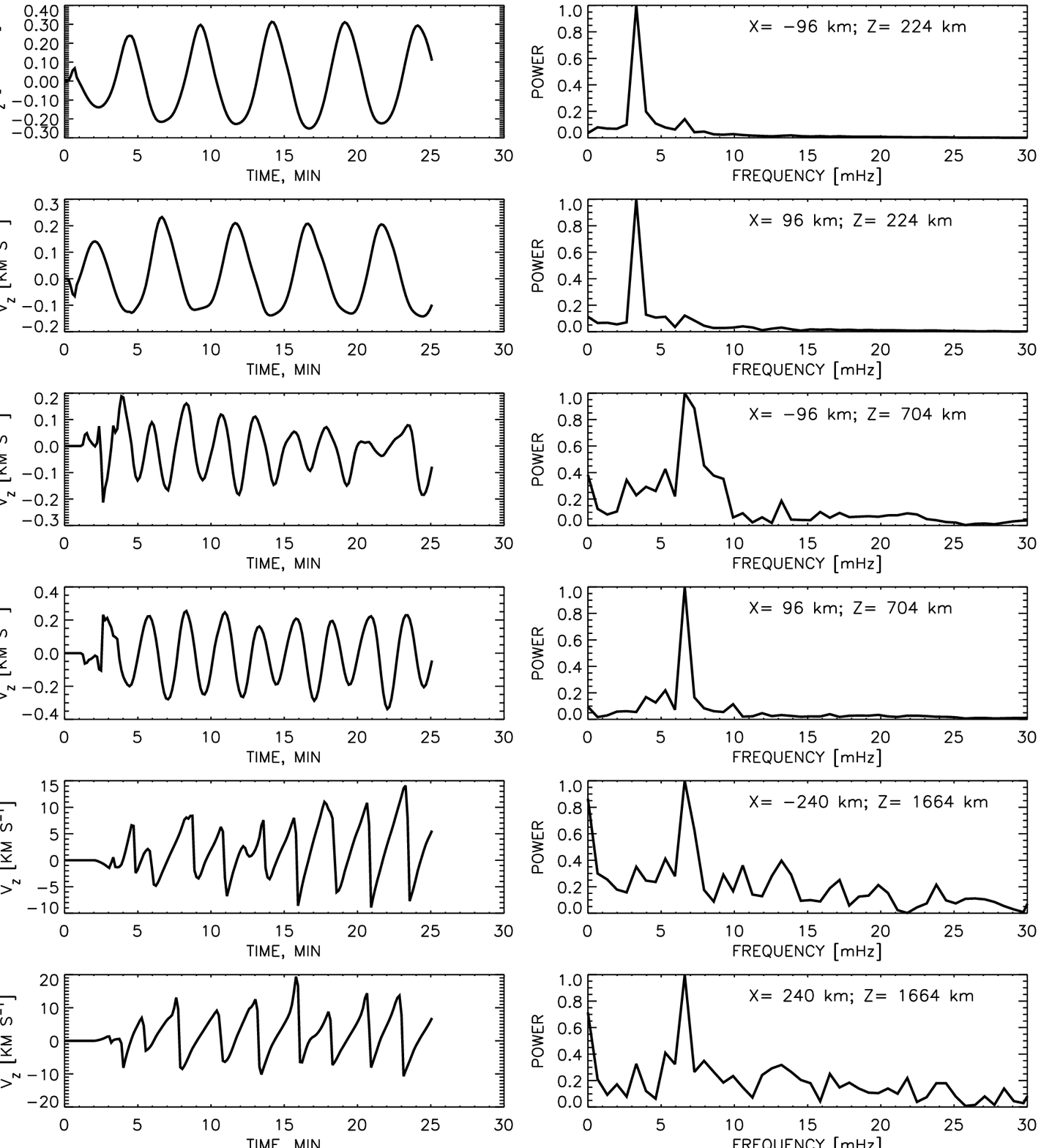}}
              \caption{Examples of the temporal evolution of the longitudinal velocity
              (left) and their corresponding power spectra (right) at
              different heights and positions inside the flux tube for the simulations
              with horizontal driving at 300 seconds. The coordinates
              are indicated on the right panels.}
\label{fig:vz300h}
\end{figure}

Figure~\ref{fig:ft300h}  shows the temporal evolution of the
transverse and longitudinal velocities over one wave period, after
the simulations reach the stationary state, and
Figure~\ref{fig:vz300h} gives the temporal variation of the
longitudinal velocity at selected points, together with their
corresponding power spectra, for the simulations with a horizontal
driver with a period of 300 seconds. The frequency of the driver
is below the cut-off frequency.
As in two previous cases, the driver excites the slow and surface
modes.
Note that, unlike the shorter-period simulations, there are
significant motions excited outside the flux tube in the
non-magnetic atmosphere as well.
The nature of these motions is partly acoustic and partly
convective since the lower layers of the photosphere are still
unstable to convection.

The slow-to-slow mode transformation is only partial, and part of
the energy of the slow mode is transferred to the fast mode above
the $v_A = c_S$ layer, producing a horizontal shaking of the tube
that propagates upwards with the Alfv\'en speed. The behavior of
the fast mode seen in the transverse velocity above $v_A=c_S$
height is similar to the three-minute period simulations, but the
left-right symmetry pattern is lost.

Most of the energy in high layers is nonetheless still carried by
the slow (acoustic) mode, giving rise to shock waves with
amplitudes of $10-15$ \kms\ above 1000 km height.
The antisymmetric nature of these shocks has disappeared. Left and
right parts of the tube are almost in phase now. However, the most
conspicuous difference lies in the period of the wave. As it
corresponds to an evanescent wave, the five-minute oscillation is
rapidly damped, and the residuals coming from non-linear effects
at 6 Hz, twice the driver frequency, already dominate at a height
of 700 km. This double frequency is clear in the temporal series
of the velocity, and in the corresponding power spectra, at this
height (two middle panels in Figure~\ref{fig:vz300h}).
From 700 km upward, the velocity increases again, due to the
density decrease, giving rise to shocks in the upper part of the
simulation domain. The main frequency is maintained around 6 mHz,
once the five-minute oscillation has been damped.
\inlinecite{Kalkofen+etal1994} and \inlinecite{Fleck+Schmitz1991}
showed that the change of the dominant period of acoustic
oscillations with height, from five to three minutes, is due to
the resonant excitation of waves at the atmospheric cut-off
frequency (\ie\ the cut-off frequency corresponding to the
temperature minimum).
However, this effect may not be dominant in our simulations since
the acoustic mode only appears above 1000 km, \ie\ well above the
temperature minimum.
At heights above 1000 km, the cut-off frequency is again around
three mHz and the resonant excitation cannot produce oscillations
at the higher frequency.
%
%
We conclude that the generation of oscillations at the first
harmonic of the driver in our simulations is mostly a non-linear
phenomenon.

As in the previous sections, the most important properties of the
wave propagation from the simulations with the horizontal driver
at 300 seconds are:

\begin{itemize}
\item Partial slow to slow (acoustic) transformation at $v_A=c_S$ height.
\item Horizontal flux tube motion at 2000 km due to the fast (magnetic)
wave.
\item Left-right shock wave pattern in phase.
\item Surface mode excitation.
\item Period of the wave not maintained with height. \\
\end{itemize}

\section{Vertical Driving at 300 Seconds}

\begin{figure}
\centerline{\includegraphics[height=0.95\textheight]{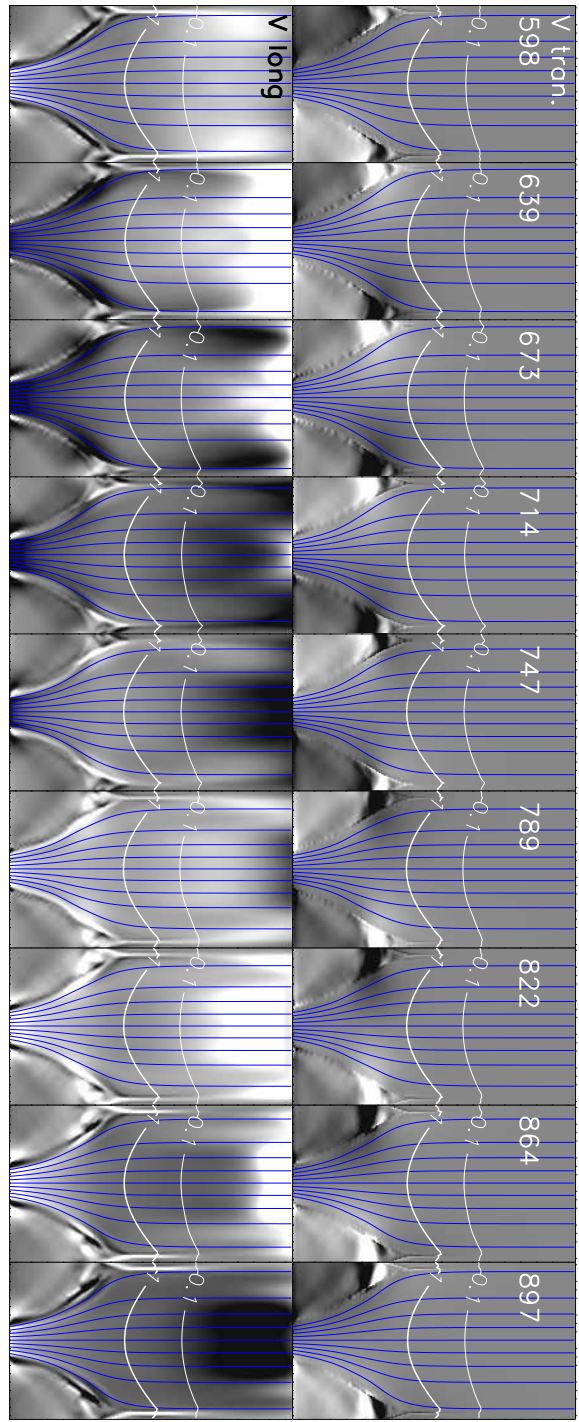}}
              \caption{Time series of snapshots of the transverse (right)
              and longitudinal (left) velocities in the simulation  with
              a vertical driver at 300 seconds. Numbers give time
              in seconds since the start of the simulation. After removing the average velocity increase with height produced
              by the density fall-off, all images have the same scale. Each snapshot includes 900
              km in horizontal and 2000 km in vertical direction.} \label{fig:ft300v}
\end{figure}

\begin{figure}
\centerline{\includegraphics[width=0.8\textwidth]{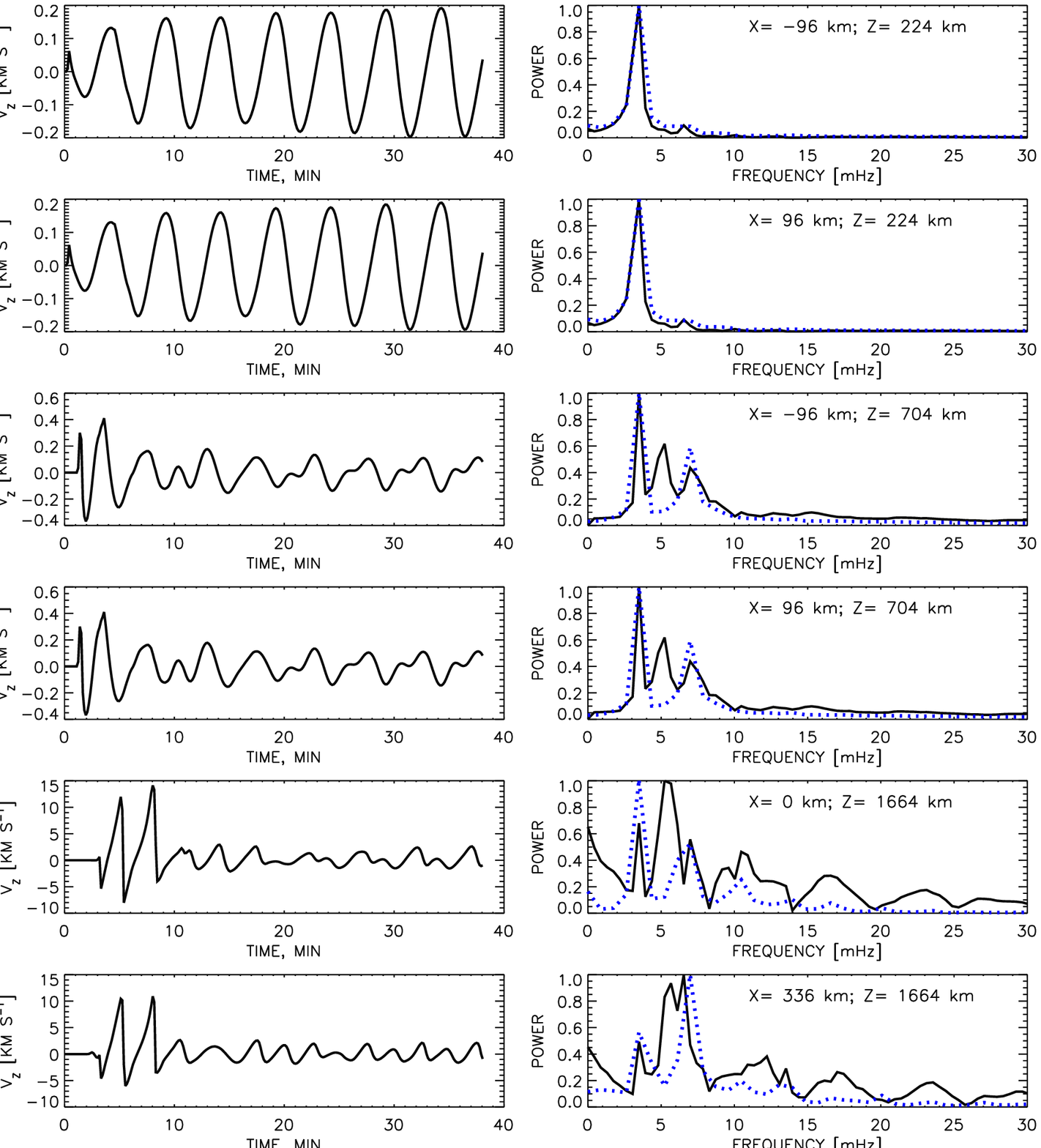}}
              \caption{Examples of the time evolution of the longitudinal velocity (left) and
              their corresponding power spectra (right) at different heights and positions inside
              the flux tube in simulations with vertical driving at 300 seconds. The coordinates are indicated on the right
              panels. The blue dotted line at the right panels
              gives the power spectra only for the last half of
              the simulations after reaching the stationary state.
              } \label{fig:vz300v}
\end{figure}

Finally, the results after exciting the flux tube with a
300-seconds vertical driver are presented. Figure~\ref{fig:ft300v}
gives some snapshots of the velocity evolution during one period
with the same format as in the previous figures.
The time series shown in the figure is taken after the simulations
reach the stationary state.
Now, the fast (acoustic) wave is excited directly by the driver
inside the flux tube in the deeper layers (see lower panels of
Figure~\ref{fig:ft300v}).
The surface wave exists as well at the magnetic/field-free
interface.
There are some acoustic disturbances that can be observed in the
field-free atmosphere surrounding the tube at the lower layers.
After some time has elapsed since the start of the simulation,
convective vortices appear in the non-magnetic atmosphere.

At the $v_A=c_S$ layer, the fast to slow transformation occurs,
keeping the acoustic nature of the wave. Almost no energy is
transferred to the fast magnetic mode in high layers.
It can be seen in the upper panels of Figure~\ref{fig:ft300v} that
the transverse velocity inside the flux tube remains close to zero
all of the time.
The magnetic field lines are compressed and expanded as the wave
passes through the atmosphere, but no horizontal motions are
observed.

The perturbation is almost perfectly symmetric at all heights with
respect to the tube axis, as demonstrated in the four upper panels
of Figure~\ref{fig:vz300v}, where the time series in symmetric
positions inside the flux tubes are given.
The most striking fact comes from the different behavior of the
velocity at points near the tube axis or at the tube boundary in
the upper heights.
At deep layers, the propagation is mainly linear showing the five-
minute sinusoidal pattern of the driver (see the upper panels in
Figure~\ref{fig:vz300v}). From there upwards, two shock waves
develop at the beginning of the series, which are followed by a
highly damped oscillation. The period of these shocks is three
minutes.
After the simulation reaches the stationary state, the dominant
periods are different at the tube axis and off axis.
For points near the axis, the five-minute oscillation does not
suffer such a large amplitude damping and is the dominate period
once the oscillation reaches a stationary state (see the power
spectra corresponding to the second half of the simulation as a
blue line in Figure~\ref{fig:vz300v}).
For off-axis points, there is a clear change with height of the
dominating frequency. The contribution of the five-minute
fundamental frequency is reduced as the wave propagates,
increasing the relative importance of the first harmonic at 6 mHz
(lower panel of Figure~\ref{fig:vz300v}).
Thus, in this simulation, 3 and 6 mHz oscillations coexist in
different parts of the tube.
In any case, the amplitudes are rather small when compared to
those derived from horizontal drivers and only weak eventual
shocks are observed with amplitudes of $2-3$ \kms.

To summarize, the most important features of the simulations with
a horizontal driver of 300 seconds are:

\begin{itemize}
\item Fast to slow transformation at $v_A=c_S$ height.
\item Weak eventual shocks
\item Period on-axis: five minutes
\item Period off-axis: three minutes
\end{itemize}

If the simulations with horizontal drivers can be understood in
terms of mode transformation, and wave amplification/damping, this
is not the case for this simulation with a 300-seconds vertical
driver. Why does the main period of the wave above the $v_A = c_S$
layer depend on the distance to the axis? Why do shocks develop at
the beginning of the series and not later? More simulations are
needed in different flux tubes, with varying initial conditions,
to understand this complex behavior.

\section{Discussion and Conclusions}

We have performed simulations of waves in flux tubes with
non-trivial magnetic field configuration. The model flux tube has
horizontal and vertical variations of magnetic field strength and
gas pressure and of the ratio between the characteristic wave
speeds $v_A$ and $c_S$. These properties have allowed us to study
wave propagation and mode transformation inside the flux tube.
Waves are excited by a photospheric driver with a period of three
and five minutes, for the first time with a magnetic field
configuration of this kind.
Different wave patterns are observed depending on the period and
on the type of the driver.
The following items summarize our findings:
\begin{itemize}
\item Horizontal motions of the flux tube at the bottom photospheric boundary
generate a slow magneto-acoustic mode inside the tube and a surface
mode at the magnetic/field free interface.
\item After the slow magneto-acoustic mode and the surface mode
reach the height where $v_A \approx c_S$, their energy is
effectively transformed into a slow acoustic mode in the high
atmosphere where $v_A > c_S$. Only a small part of the driver
energy is returned back to the photosphere by the fast
magneto-acoustic mode.
\item The slow acoustic mode propagates vertically along the
magnetic field lines in the atmosphere where $v_A > c_S$, forms
shock waves above 1000 km with amplitudes of $5-15$ \kms\ and
remains always within the same flux tube. Thus, it can deposit
effectively most part of the energy of the driver into the
chromosphere.
\item If the frequency of the horizontal driver is above the acoustic
cut-off, non-linear wave propagation in the tube occurs with the
dominant period of the driver at all heights. Non-linear effects
produce higher harmonics, but their amplitude is small.
\item If the frequency of the horizontal driver is below the acoustic cut-off,
the dominant period of the longitudinal velocity changes with
height from five to three minutes due to non-linear generation of
the higher harmonics.
%
\item The five-minute vertical perturbation of the flux tube
at the bottom photospheric boundary generates a fast acoustic mode
that propagates upward through the transformation $v_A = c_S$
layer, without changing its nature and forming only eventual
shocks in the chromosphere.
\item After reaching the stationary state in the simulations with
a vertical driver at 300 seconds, both three and five minute
oscillations co-exist inside the flux tube at chromospheric
heights. The dominant period at the axis is five minutes and the
off-axis dominant period is three minutes.
\end{itemize}

Our simulations suggest once again that the properties of waves
observed in magnetic structures are the direct consequence of
their magnetic configuration, temperature, magnetic field
strength, and the height where the mode transformation occurs.
When analyzing observations, it is important to know whether they
correspond to the level below or above the layer $v_A = c_S$.
If the motions that excite oscillations inside flux tubes are
purely horizontal, no correlation may be observed between velocity
variations measured in the photosphere and the chromosphere.
This is because only an acoustic mode can be detected in
observations (at least at disc center) since it produces
significant vertical velocities, unlike the slow magnetic mode
existing in the photosphere.
The acoustic mode is only generated above $700-1000$ km in our
simulations with a horizontal driver.
In contrast, if the driver that excites oscillations has a
vertical component, the acoustic mode is generated already in the
photosphere.
In this case, the oscillations in vertical velocity are coherent
at photospheric and chromospheric heights.

In general, both simulations with horizontal and vertical drivers
with 300-second periodicity show the change of the dominant wave
period with height.
This property is different from observations of plage and network
regions where the five minute periodicity is preserved also in the
chromosphere (\opencite{Lites+Rutten+Kalkofen1993},
\opencite{Krijer+etal2001}, \opencite{DePontieu+etal2003},
\opencite{Centeno+etal2006b}, \opencite{Bloomfield+etal2006},
\opencite{Veccio+etal2007}).
While our study may help to identify the wave modes observed in
small-scale magnetic structures, further work is needed in order
to explain the behavior of the oscillatory spectra with height.

\acknowledgements{The authors are grateful for the anonymous
referee for his suggestions that helped to improve the manuscript.
Financial support by the European Commission through the SOLAIRE
Network (MTRN-CT-2006-035484) and by the Spanish Ministery of
Education through projects AYA2007-66502 and AYA2007-63881 is
gratefully acknowledged.}


\end{article}

\begin{thebibliography}{53}
\ifx \bisbn   \undefined \def \bisbn  #1{ISBN #1}   \fi
\ifx \binits  \undefined \def \binits#1{#1} \fi
\ifx \bauthor  \undefined \def \bauthor#1{#1} \fi
\ifx \batitle  \undefined \def \batitle#1{#1} \fi
\ifx \bjtitle  \undefined \def \bjtitle#1{#1} \fi
\ifx \bvolume  \undefined \def \bvolume#1{#1} \fi
\ifx \byear  \undefined \def \byear#1{#1} \fi
\ifx \bissue  \undefined \def \bissue#1{#1} \fi
\ifx \bfpage  \undefined \def \bfpage#1{#1} \fi
\ifx \blpage  \undefined \def \blpage #1{#1} \fi
\ifx \burl  \undefined \def \burl#1{#1} \fi
\ifx \binterref  \undefined \def \binterref#1{#1} \fi
\ifx \betal  \undefined \def \betal#1{#1} \fi
\ifx \binstitute  \undefined \def \binstitute#1{#1} \fi
\ifx \bctitle  \undefined \def \bctitle#1{#1} \fi
\ifx \beditor  \undefined \def \beditor#1{#1} \fi
\ifx \bpublisher  \undefined \def \bpublisher#1{#1} \fi
\ifx \bbtitle  \undefined \def \bbtitle#1{#1} \fi
\ifx \bedition  \undefined \def \bedition#1{#1} \fi
\ifx \bseriesno  \undefined \def \bseriesno#1{#1} \fi
\ifx \blocation  \undefined \def \blocation#1{#1} \fi
\ifx \bsertitle  \undefined \def \bsertitle#1{#1} \fi
\ifx \bsnm \undefined \def \bsnm#1{#1} \fi
\ifx \bsuffix \undefined \def \bsuffix#1{#1} \fi
\ifx \bparticle \undefined \def \bparticle#1{#1} \fi
\ifx \barticle \undefined \def \barticle#1{#1} \fi
\ifx \botherref \undefined \def \botherref #1{#1} \fi
\ifx \url \undefined \def \url#1{\textsf{#1}} \fi
\ifx \bchapter \undefined \def \bchapter#1{#1} \fi
\ifx \bbook \undefined \def \bbook#1{#1} \fi
\ifx \bcomment \undefined \def \bcomment#1{#1} \fi
\ifx \oauthor \undefined \def \oauthor#1{#1} \fi
\ifx \citeauthoryear \undefined \def \citeauthoryear#1{#1} \fi
\def \endbibitem {}


\bibitem[\protect\citeauthoryear{{\mbox{Bellot Rubio}} {\it
  et~al.}}{2000}]{BellotRubio+Collados+RuisCobo+RodriguezHidalgo2000}
\begin{barticle}
\bauthor{\bsnm{\mbox{Bellot Rubio}},~\binits{L.R.}},
  \bauthor{\bsnm{Collados},~\binits{M.}}, \bauthor{\bsnm{\mbox{Ruiz
  Cobo}},~\binits{B.}}, \bauthor{\bsnm{\mbox{Rodr\'{\i}guez
  Hidalgo}},~\binits{I.}}:
\byear{2000}, \batitle{Oscillations in the photosphere of a sunspot umbra from
  the inversion of infrared stokes profiles}. \textit{\bjtitle{Astrophys. J.}}
  \textbf{\bvolume{534}}, \bfpage{989}\,--\,\blpage{996}.
\end{barticle}
\endbibitem


\bibitem[\protect\citeauthoryear{Berenger}{1994}]{Berenger1994}
\begin{barticle}
\bauthor{\bsnm{Berenger},~\binits{J.P.}}:
\byear{1994}, \batitle{A perfectly mached layer for the
absorption of electromagnetic waves}.  \textit{\bjtitle{J.\ Comp.\ Phys.}} 
\textbf{\bvolume{114}}, \bfpage{185}\,--\,\blpage{200}.
\end{barticle}
\endbibitem

\bibitem[\protect\citeauthoryear{Berger, \mbox{Rouppe van der Voort}, and
  L{\"o}fdahl}{2007}]{Berger+etal2007}
\begin{barticle}
\bauthor{\bsnm{Berger},~\binits{T.E.}}, \bauthor{\bsnm{\mbox{Rouppe van der
  Voort}},~\binits{L.}}, \bauthor{\bsnm{L{\"o}fdahl},~\binits{M.}}:
\byear{2007}, \batitle{Contrast analysis of solar faculae and magnetic bright
  points}. \textit{\bjtitle{Astrophys. J.}} \textbf{\bvolume{661}},
  \bfpage{1272}\,--\,\blpage{1288}.
\end{barticle}
\endbibitem

\bibitem[\protect\citeauthoryear{Bloomfield, Lagg, and
  Solanki}{2007}]{Bloomfield+etal2007b}
\begin{botherref}
\oauthor{\bsnm{Bloomfield},~\binits{D.S.}}, \oauthor{\bsnm{Lagg},~\binits{A.}},
  \oauthor{\bsnm{Solanki},~\binits{S.K.}}:
2007, Observations of running waves in a sunspot chromosphere. In: Heinzel, P.,
  Dorotovic, I., Rutten, R.J. (eds.) \textit{The Physics of Chromospheric
  Plasmas},  Astron. Soc. Pac. Conference Series, San Francisco, \textbf{368}, 239.
\end{botherref}
\endbibitem

\bibitem[\protect\citeauthoryear{{Bloomfield} {\it
  et~al.}}{2006}]{Bloomfield+etal2006}
\begin{barticle}
\bauthor{\bsnm{Bloomfield},~\binits{D.S.}},
  \bauthor{\bsnm{McAteer},~\binits{R.T.J.}},
  \bauthor{\bsnm{Mathioudakis},~\binits{M.}},
  \bauthor{\bsnm{Keenan},~\binits{F.P.}}:
\byear{2006}, \batitle{The influence of magnetic field on oscillations in the
  solar chromosphere}. \textit{\bjtitle{Astrophys. J.}} \textbf{\bvolume{652}},
  \bfpage{812}\,--\,\blpage{819}.
\end{barticle}
\endbibitem

\bibitem[\protect\citeauthoryear{{Bogdan} {\it et~al.}}{2003}]{Bogdan+etal2003}
\begin{barticle}
\bauthor{\bsnm{Bogdan},~\binits{T.J.}}, \bauthor{\bsnm{Carlsson},~\binits{M.}},
  \bauthor{\bsnm{Hansteen},~\binits{V.}},
  \bauthor{\bsnm{McMurry},~\binits{A.}},
  \bauthor{\bsnm{Rosenthal},~\binits{C.S.}},
  \bauthor{\bsnm{Johnson},~\binits{M.}},
  \bauthor{\bsnm{Petty-Powell},~\binits{S.}},
  \bauthor{\bsnm{Zita},~\binits{E.J.}}, \bauthor{\bsnm{Stein},~\binits{R.F.}},
  \bauthor{\bsnm{McIntosh},~\binits{S.W.}},
  \bauthor{\bsnm{Nordlund},~\binits{A.}}:
\byear{2003}, \batitle{Waves in the magnetized solar atmosphere. ii. waves from
  localized sources in magnetic flux concentrations}. \textit{\bjtitle{Astrophys. J.}}
  \textbf{\bvolume{599}}, \bfpage{626}\,--\,\blpage{660}.
\end{barticle}
\endbibitem

\bibitem[\protect\citeauthoryear{Bogdan and Judge}{2006}]{Bogdan+Judge2006}
\begin{botherref}
\oauthor{\bsnm{Bogdan},~\binits{T.J.}}, \oauthor{\bsnm{Judge},~\binits{P.G.}}:
2006, Observational aspects of sunspot oscillations. In: \textit{MHD wave and
  oscillations in the Solar Plasma}, 
  Phil. Trans. Royal. Soc., \textbf{364, Issue 1839}, 313\,--\,331.
\end{botherref}
\endbibitem

\bibitem[\protect\citeauthoryear{Braun and Lindsey}{1999}]{Braun+Linsday1999}
\begin{barticle}
\bauthor{\bsnm{Braun},~\binits{D.C.}}, \bauthor{\bsnm{Lindsey},~\binits{C.}}:
\byear{1999}, \batitle{Helioseismic images of an active region complex}.
  \textit{\bjtitle{Astrophys. J.}} \textbf{\bvolume{513}},
  \bfpage{L79}\,--\,\blpage{L82}.
\end{barticle}
\endbibitem

\bibitem[\protect\citeauthoryear{{Brynildsen} {\it
  et~al.}}{2002}]{Brynildsen+etal2002}
\begin{barticle}
\bauthor{\bsnm{Brynildsen},~\binits{N.}}, \bauthor{\bsnm{Maltby},~\binits{P.}},
  \bauthor{\bsnm{Fredvik},~\binits{T.}},
  \bauthor{\bsnm{Kjeldseth-Moe},~\binits{O.}}:
\byear{2002}, \batitle{Oscillations above sunspots}. \textit{\bjtitle{Solar
  Phys.}} \textbf{\bvolume{207}}, \bfpage{259}\,--\,\blpage{290}.
\end{barticle}
\endbibitem

\bibitem[\protect\citeauthoryear{{Brynildsen} {\it
  et~al.}}{2000}]{Brynildsen+etal2000}
\begin{barticle}
\bauthor{\bsnm{Brynildsen},~\binits{N.}}, \bauthor{\bsnm{Maltby},~\binits{P.}},
  \bauthor{\bsnm{Leifsen},~\binits{T.}},
  \bauthor{\bsnm{Kjeldseth-Moe},~\binits{O.}},
  \bauthor{\bsnm{Wilhelm},~\binits{K.}}:
\byear{2000}, \batitle{Observations of sunspot transition region oscillations}.
  \textit{\bjtitle{Solar Phys.}} \textbf{\bvolume{191}},
  \bfpage{129}\,--\,\blpage{159}.
\end{barticle}
\endbibitem

\bibitem[\protect\citeauthoryear{Cally}{2006}]{Cally2006}
\begin{barticle}
\bauthor{\bsnm{Cally},~\binits{P.}}:
\byear{2006}, \batitle{Dispersion relations, rays and ray splitting in magnetohelioseismology}.
\textit{\bjtitle{Royal Soc. London Trans. Series A}} 
\textbf{\bvolume{364}},
  \bfpage{333}\,--\,\blpage{349}.
\end{barticle}
\endbibitem


\bibitem[\protect\citeauthoryear{Cally}{2007}]{Cally2007}
\begin{barticle}
\bauthor{\bsnm{Cally},~\binits{P.}}:
\byear{2006}, \batitle{What to look for in the seismology of solar active regions}.
\textit{\bjtitle{Astronomische Nachrichten}} 
\textbf{\bvolume{328}},
  \bfpage{286}.
\end{barticle}
\endbibitem



\bibitem[\protect\citeauthoryear{Carlsson and Stein}{1997}]{Carlsson+Stein1997}
\begin{barticle}
\bauthor{\bsnm{Carlsson},~\binits{M.}}, \bauthor{\bsnm{Stein},~\binits{R.F.}}:
\byear{1997}, \batitle{Formation of Solar Calcium H and K Bright Grains}. \textit{\bjtitle{Astrophys. J.}}
  \textbf{\bvolume{481}}, \bfpage{500}.
\end{barticle}
\endbibitem




\bibitem[\protect\citeauthoryear{Centeno, Collados, and \mbox{Trujillo
  Bueno}}{2006}]{Centeno+etal2006b}
\begin{botherref}
\oauthor{\bsnm{Centeno},~\binits{R.}}, \oauthor{\bsnm{Collados},~\binits{M.}},
  \oauthor{\bsnm{\mbox{Trujillo Bueno}},~\binits{J.}}:
2006, Oscillations and wave propagation in different solar magnetic features.
  In: Casini, R., Lites, B.W. (eds.) \textit{Solar Polarization 4},
   Astron. Soc. Pac. Conference Series, San Francisco, \textbf{358}, 465.
\end{botherref}
\endbibitem

\bibitem[\protect\citeauthoryear{Centeno, Collados, and \mbox{Trujillo
  Bueno}}{2006}]{Centeno+etal2006a}
\begin{barticle}
\bauthor{\bsnm{Centeno},~\binits{R.}}, \bauthor{\bsnm{Co\-lla\-dos},~\binits{M.}},
  \bauthor{\bsnm{\mbox{Trujillo Bueno}},~\binits{J.}}:
\byear{2006}, \batitle{Spectropolarimetric investigation of the propagation of
  magnetoacoustic waves and shock formation in sunspot atmospheres}.
  \textit{\bjtitle{Astrophys. J.}} \textbf{\bvolume{640}},
  \bfpage{1153}\,--\,\blpage{1162}.
\end{barticle}
\endbibitem

\bibitem[\protect\citeauthoryear{Christopoulou, Georgakilas, and
  Koutchmy}{2000}]{Christopoulou+etal2000}
\begin{barticle}
\bauthor{\bsnm{Christopoulou},~\binits{E.B.}},
  \bauthor{\bsnm{Georgakilas},~\binits{A.A.}},
  \bauthor{\bsnm{Koutchmy},~\binits{S.}}:
\byear{2000}, \batitle{Oscillations and running waves observed in sunspots}.
  \textit{\bjtitle{Astron. Astrophys.}} \textbf{\bvolume{354}},
  \bfpage{305}\,--\,\blpage{314}.
\end{barticle}
\endbibitem

\bibitem[\protect\citeauthoryear{Christopoulou, Georgakilas, and
  Koutchmy}{2001}]{Christopoulou+etal2001}
\begin{barticle}
\bauthor{\bsnm{Christopoulou},~\binits{E.B.}},
  \bauthor{\bsnm{Georgakilas},~\binits{A.A.}},
  \bauthor{\bsnm{Koutchmy},~\binits{S.}}:
\byear{2001}, \batitle{Oscillations and running waves observed in sunspots.
  iii. multilayer study}. \textit{\bjtitle{Astron. Astrophys.}} \textbf{\bvolume{375}},
  \bfpage{617}\,--\,\blpage{628}.
\end{barticle}
\endbibitem

\bibitem[\protect\citeauthoryear{{Collados} {\it et~al.}}{2001}]{Collados2001}
\begin{botherref}
\oauthor{\bsnm{Co\-lla\-dos},~\binits{M.}}, \oauthor{\bsnm{\mbox{Trujillo
  Bueno}},~\binits{J.}}, \oauthor{\bsnm{\mbox{Bellot Rubio}},~\binits{L.R.}},
  \oauthor{\bsnm{Socas-Navarro},~\binits{H.}}:
2001, In: Ballester, J.L., Roberts, B. (eds.) \textit{INTAS workshop on MHD
  waves in astrophysical plasmas}. Universitat de les Illes Balears,
  151\,--\,154.
\end{botherref}
\endbibitem

\bibitem[\protect\citeauthoryear{{De Moortel} {\it
  et~al.}}{2002}]{DeMoortel+etal2002}
\begin{barticle}
\bauthor{\bsnm{De Moortel},~\binits{I.}}, \bauthor{\bsnm{Ireland},~\binits{J.}},
  \bauthor{\bsnm{Hood},~\binits{A.W.}}, \bauthor{\bsnm{Walsh},~\binits{R.W.}}:
\byear{2002}, \batitle{The detection of 3 \& 5 min period oscillations in
  coronal loops}. \textit{\bjtitle{Astron. Astrophys.}} \textbf{\bvolume{387}},
  \bfpage{L13}\,--\,\blpage{L16}.
\end{barticle}
\endbibitem

\bibitem[\protect\citeauthoryear{\mbox{De Pontieu}, Erdelyi, and
  de~Wijn}{2003}]{DePontieu+etal2003}
\begin{barticle}
\bauthor{\bsnm{\mbox{De Pontieu}},~\binits{B.}},
  \bauthor{\bsnm{Erdelyi},~\binits{R.}},
  \bauthor{\bparticle{de~}\bsnm{Wijn},~\binits{A.G.}}:
\byear{2003}, \batitle{Intensity oscillations in the upper transition region
  above active region plage}. \textit{\bjtitle{Astrophys. J.}} \textbf{\bvolume{595}},
  \bfpage{L63}\,--\,\blpage{L66}.
\end{barticle}
\endbibitem

\bibitem[\protect\citeauthoryear{\mbox{De Pontieu}, Erdelyi, and
  Stewart}{2004}]{DePontieu+etal2004}
\begin{barticle}
\bauthor{\bsnm{\mbox{De Pontieu}},~\binits{B.}},
  \bauthor{\bsnm{Erdelyi},~\binits{R.J.}},
  \bauthor{\bsnm{Stewart},~\binits{P.}}:
\byear{2004}, \batitle{Solar chromospheric spicules from the leakage of
  photospheric oscillations and flows}. \textit{\bjtitle{Nature}}
  \textbf{\bvolume{430}}, \bfpage{536}\,--\,\blpage{539}.
\end{barticle}
\endbibitem

\bibitem[\protect\citeauthoryear{Fleck and Schmitz}{1991}]{Fleck+Schmitz1991}
\begin{barticle}
\bauthor{\bsnm{Fleck},~\binits{B.}}, \bauthor{\bsnm{Schmitz},~\binits{F.}}:
\byear{1991}, \batitle{The 3-min oscillations of the solar chromosphere - a
  basic physical effect?} \textit{\bjtitle{Astron. Astrophys.}} \textbf{\bvolume{250}},
  \bfpage{235}\,--\,\blpage{244}.
\end{barticle}
\endbibitem

\bibitem[\protect\citeauthoryear{Fleck and Schmitz}{1993}]{Fleck+Schmitz1993}
\begin{barticle}
\bauthor{\bsnm{Fleck},~\binits{B.}}, \bauthor{\bsnm{Schmitz},~\binits{F.}}:
\byear{1993}, \batitle{On the interactions of hydrodynamic shock waves in
  stellar atmospheres}. \textit{\bjtitle{Astron. Astrophys.}} \textbf{\bvolume{273}}, 671.
\end{barticle}
\endbibitem

\bibitem[\protect\citeauthoryear{Gurman and
  Leibacher}{1984}]{Gurman+Leibacher1984}
\begin{barticle}
\bauthor{\bsnm{Gurman},~\binits{J.B.}},
  \bauthor{\bsnm{Leibacher},~\binits{J.W.}}:
\byear{1984}, \batitle{Linear models of acoustic waves in sunspot umbrae}.
  \textit{\bjtitle{Astrophys. J.}} \textbf{\bvolume{283}},
  \bfpage{859}\,--\,\blpage{869}.
\end{barticle}
\endbibitem

\bibitem[\protect\citeauthoryear{{Hasan} {\it et~al.}}{2003}]{Hasan+etal2003}
\begin{barticle}
\bauthor{\bsnm{Hasan},~\binits{S.S.}}, \bauthor{\bsnm{Kalkofen},~\binits{W.}},
  \bauthor{\bsnm{\mbox{van Ballegooijen}},~\binits{A.A.}},
  \bauthor{\bsnm{Ulmschneider},~\binits{P.}}:
\byear{2003}, \batitle{Kink and longitudinal oscillations in the magnetic
  network of the sun: Nonlinear effects and mode transformation}.
  \textit{\bjtitle{Astrophys. J.}} \textbf{\bvolume{585}},
  \bfpage{1138}\,--\,\blpage{1146}.
\end{barticle}
\endbibitem

\bibitem[\protect\citeauthoryear{{Hasan} {\it et~al.}}{2005}]{Hasan+etal2005}
\begin{barticle}
\bauthor{\bsnm{Hasan},~\binits{S.S.}}, \bauthor{\bsnm{\mbox{van
  Ballegooijen}},~\binits{A.A.}}, \bauthor{\bsnm{Kalkofen},~\binits{W.}},
  \bauthor{\bsnm{Steiner},~\binits{O.}}:
\byear{2005}, \batitle{Dynamics of solar magnetic network: two-dimensional mhd
  simulations}. \textit{\bjtitle{Astrophys. J.}} \textbf{\bvolume{631}},
  \bfpage{1270}\,--\,\blpage{1280}.
\end{barticle}
\endbibitem

\bibitem[\protect\citeauthoryear{Hasan and
  Ulmschneider}{2004}]{Hasan+Ulmschneider2004}
\begin{barticle}
\bauthor{\bsnm{Hasan},~\binits{S.S.}},
  \bauthor{\bsnm{Ulmschneider},~\binits{P.}}:
\byear{2004}, \batitle{Dynamisc and heating of the magnetic network on the sun.
  efficiency of mode transformation}. \textit{\bjtitle{Astron. Astrophys.}}
  \textbf{\bvolume{422}}, \bfpage{1085}\,--\,\blpage{1091}.
\end{barticle}
\endbibitem

\bibitem[\protect\citeauthoryear{Judge, Tarbell, and
  Wilhelm}{2001}]{Judge+etal2001}
\begin{barticle}
\bauthor{\bsnm{Judge},~\binits{P.G.}}, \bauthor{\bsnm{Tarbell},~\binits{T.D.}},
  \bauthor{\bsnm{Wilhelm},~\binits{K.}}:
\byear{2001}, \batitle{A study of chromospheric oscillations using the soho and
  trace spacecraft}. \textit{\bjtitle{Astrophys. J.}} \textbf{\bvolume{554}},
  \bfpage{424}\,--\,\blpage{444}.
\end{barticle}
\endbibitem

\bibitem[\protect\citeauthoryear{{Kalkofen} {\it
  et~al.}}{1994}]{Kalkofen+etal1994}
\begin{barticle}
\bauthor{\bsnm{Kalkofen},~\binits{W.}}, \bauthor{\bsnm{Rossi},~\binits{P.}},
  \bauthor{\bsnm{Bodo},~\binits{G.}}, \bauthor{\bsnm{Massaglia},~\binits{S.}}:
\byear{1994}, \batitle{Propagation of acoustic waves in a stratified
  atmosphere}. \textit{\bjtitle{Astron. Astrophys.}} \textbf{\bvolume{284}},
  \bfpage{976}\,--\,\blpage{984}.
\end{barticle}
\endbibitem

\bibitem[\protect\citeauthoryear{Khomenko and
  Collados}{2006}]{Khomenko+Collados2006}
\begin{barticle}
\bauthor{\bsnm{Khomenko},~\binits{E.}}, \bauthor{\bsnm{Collados},~\binits{M.}}:
\byear{2006}, \batitle{Numerical modeling of magnetohydrodynamic wave
  propagation and refraction in sunspots}. \textit{\bjtitle{Astrophys. J.}}
  \textbf{\bvolume{653}}, \bfpage{739}\,--\,\blpage{755}.
\end{barticle}
\endbibitem

\bibitem[\protect\citeauthoryear{Khomenko, Collados, and \mbox{Bellot
  Rubio}}{2003}]{Khomenko+Collados+BellotRubio2003}
\begin{barticle}
\bauthor{\bsnm{Khomenko},~\binits{E.V.}},
  \bauthor{\bsnm{Collados},~\binits{M.}}, \bauthor{\bsnm{\mbox{Bellot
  Rubio}},~\binits{L.R.}}:
\byear{2003}, \batitle{Magnetoacoustic waves in sunspots}.
  \textit{\bjtitle{Astrophys. J.}} \textbf{\bvolume{588}},
  \bfpage{606}\,--\,\blpage{619}.
\end{barticle}
\endbibitem

\bibitem[\protect\citeauthoryear{Korn and Korn}{2000}]{Korn}
\begin{botherref}
\oauthor{\bsnm{Korn},~\binits{G.A.}}, \oauthor{\bsnm{Korn},~\binits{T.M.}}:
2000, Mathematical Handbook for Scientists and Engineers. Dover.
\end{botherref}
\endbibitem

\bibitem[\protect\citeauthoryear{{Krijger} {\it
  et~al.}}{2001}]{Krijer+etal2001}
\begin{barticle}
\bauthor{\bsnm{Krijger},~\binits{J.M.}},
  \bauthor{\bsnm{Rutten},~\binits{R.J.}},
  \bauthor{\bsnm{Lites},~\binits{B.W.}}, \bauthor{\bsnm{Straus},~\binits{T.}},
  \bauthor{\bsnm{Shine},~\binits{R.A.}},
  \bauthor{\bsnm{Tarbell},~\binits{T.D.}}:
\byear{2001}, \batitle{Dynamics of the solar chromosphere. iii. ultraviolet
  brightness oscillations from trace}. \textit{\bjtitle{Astron. Astrophys.}}
  \textbf{\bvolume{379}}, \bfpage{1052}\,--\,\blpage{1082}.
\end{barticle}
\endbibitem

\bibitem[\protect\citeauthoryear{Lites}{1984}]{Lites1984}
\begin{barticle}
\bauthor{\bsnm{Lites},~\binits{B.W.}}:
\byear{1984}, \batitle{Photoelectric observations of chromospheric sunspot
  oscillations. II - propagation characteristics}. \textit{\bjtitle{Astrophys. J.}}
  \textbf{\bvolume{277}}, \bfpage{874}\,--\,\blpage{888}.
\end{barticle}
\endbibitem

\bibitem[\protect\citeauthoryear{Lites}{1986}]{Lites1986}
\begin{barticle}
\bauthor{\bsnm{Lites},~\binits{B.W.}}:
\byear{1986}, \batitle{Photoelectric observations of chromospheric umbral
  oscillations. iv. the caii h line and hei 1830.} \textit{\bjtitle{Astrophys. J.}}
  \textbf{\bvolume{301}}, \bfpage{1005}\,--\,\blpage{1017}.
\end{barticle}
\endbibitem

\bibitem[\protect\citeauthoryear{Lites}{1988}]{Lites1988}
\begin{barticle}
\bauthor{\bsnm{Lites},~\binits{B.W.}}:
\byear{1988}, \batitle{Photoelectric observations of chromospheric umbral
  oscillations. v - penumbral oscillations}. \textit{\bjtitle{Astrophys. J.}}
  \textbf{\bvolume{334}}, \bfpage{1054}\,--\,\blpage{1065}.
\end{barticle}
\endbibitem

\bibitem[\protect\citeauthoryear{Lites, Rutten, and
  Kalkofen}{1993}]{Lites+Rutten+Kalkofen1993}
\begin{barticle}
\bauthor{\bsnm{Lites},~\binits{B.W.}}, \bauthor{\bsnm{Rutten},~\binits{R.J.}},
  \bauthor{\bsnm{Kalkofen},~\binits{W.}}:
\byear{1993}, \batitle{Dynamics of the solar chromosphere. I - long-period
  network oscillations}. \textit{\bjtitle{Astrophys. J.}} \textbf{\bvolume{414}},
  \bfpage{345}\,--\,\blpage{356}.
\end{barticle}
\endbibitem

\bibitem[\protect\citeauthoryear{{Lites} {\it
  et~al.}}{1998}]{Lites+Thomas+Bogdan+Cally1998}
\begin{barticle}
\bauthor{\bsnm{Lites},~\binits{B.W.}}, \bauthor{\bsnm{Thomas},~\binits{J.H.}},
  \bauthor{\bsnm{Bogdan},~\binits{T.J.}},
  \bauthor{\bsnm{Cally},~\binits{P.S.}}:
\byear{1998}, \batitle{Velocity and magnetic field fluctuations in the
  photosphere of a sunspot}. \textit{\bjtitle{Astrophys. J.}} \textbf{\bvolume{497}},
  \bfpage{464}\,--\,\blpage{482}.
\end{barticle}
\endbibitem

\bibitem[\protect\citeauthoryear{{Maltby} {\it et~al.}}{1999}]{Maltby+etal1999}
\begin{barticle}
\bauthor{\bsnm{Maltby},~\binits{P.}}, \bauthor{\bsnm{Brynildsen},~\binits{N.}},
  \bauthor{\bsnm{Fredvik},~\binits{T.}},
  \bauthor{\bsnm{Kjeldseth-Moe},~\binits{O.}},
  \bauthor{\bsnm{Wilhelm},~\binits{K.}}:
\byear{1999}, \batitle{On the sunspot transition region}.
  \textit{\bjtitle{Solar Phys.}} \textbf{\bvolume{190}},
  \bfpage{437}\,--\,\blpage{458}.
\end{barticle}
\endbibitem

\bibitem[\protect\citeauthoryear{{Maltby} {\it et~al.}}{2001}]{Maltby+etal2001}
\begin{barticle}
\bauthor{\bsnm{Maltby},~\binits{P.}}, \bauthor{\bsnm{Brynildsen},~\binits{N.}},
  \bauthor{\bsnm{Kjeldseth-Moe},~\binits{O.}},
  \bauthor{\bsnm{Wilhelm},~\binits{K.}}:
\byear{2001}, \batitle{Plumes and oscillations in the sunspot transition
  region}. \textit{\bjtitle{Astron. Astrophys.}} \textbf{\bvolume{373}},
  \bfpage{L1}\,--\,\blpage{L4}.
\end{barticle}
\endbibitem

\bibitem[\protect\citeauthoryear{Marsh and Walsh}{2005}]{Marsh+Walsh2005}
\begin{botherref}
\oauthor{\bsnm{Marsh},~\binits{M.S.}}, \oauthor{\bsnm{Walsh},~\binits{R.W.}}:
2005, Observed wave propagtion along the sunspot magnetic field through the
  chromosphere, transition region and corona. In: \textit{Chromospheric and
  Coronal Magnetic fields},  ESA SP, Noordwijk. \textbf{596}, \bfpage{75.1}.
\end{botherref}
\endbibitem

\bibitem[\protect\citeauthoryear{Marsh and Walsh}{2006}]{Marsh+Walsh2006}
\begin{barticle}
\bauthor{\bsnm{Marsh},~\binits{M.S.}}, \bauthor{\bsnm{Walsh},~\binits{R.W.}}:
\byear{2006}, \batitle{p-mode propagation through the transition region into
  the solar corona. i. observations}. \textit{\bjtitle{Astrophys. J.}}
  \textbf{\bvolume{643}}, \bfpage{540}\,--\,\blpage{548}.
\end{barticle}
\endbibitem



\bibitem[\protect\citeauthoryear{Pneuman, Solanki, and
  Stenflo}{1986}]{Pneuman+etal1986}
\begin{barticle}
\bauthor{\bsnm{Pneuman},~\binits{G.W.}},
  \bauthor{\bsnm{Solanki},~\binits{S.K.}},
  \bauthor{\bsnm{Stenflo},~\binits{J.O.}}:
\byear{1986}, \batitle{Structure and merging of solar magnetic fluxtubes}.
  \textit{\bjtitle{Astron. Astrophys.}} \textbf{\bvolume{154}},
  \bfpage{231}\,--\,\blpage{242}.
\end{barticle}
\endbibitem

\bibitem[\protect\citeauthoryear{Roberts}{1981}]{Roberts1981}
\begin{barticle}
\bauthor{\bsnm{Roberts},~\binits{B.}}:
\byear{1981}, \batitle{Wave propagation in a magnetically structured
  atmosphere. i - surface waves at a magnetic interface.}
  \textit{\bjtitle{Solar Phys.}} \textbf{\bvolume{69}},
  \bfpage{27}\,--\,\blpage{38}.
\end{barticle}
\endbibitem

\bibitem[\protect\citeauthoryear{Roberts}{1983}]{Roberts1983}
\begin{barticle}
\bauthor{\bsnm{Roberts},~\binits{B.}}:
\byear{1983}, \batitle{Wave propagation in intense flux tubes.}
  \textit{\bjtitle{Solar Phys.}} \textbf{\bvolume{87}},
  \bfpage{77}.
\end{barticle}
\endbibitem


\bibitem[\protect\citeauthoryear{{Rosenthal} {\it
  et~al.}}{2002}]{Rosenthal+etal2002}
\begin{barticle}
\bauthor{\bsnm{Rosenthal},~\binits{C.S.}},
  \bauthor{\bsnm{Bogdan},~\binits{T.J.}},
  \bauthor{\bsnm{Carlsson},~\binits{M.}},
  \bauthor{\bsnm{Dorch},~\binits{S.B.F.}},
  \bauthor{\bsnm{Hansteen},~\binits{V.}},
  \bauthor{\bsnm{McIntosh},~\binits{S.W.}},
  \bauthor{\bsnm{McMurry},~\binits{A.}},
  \bauthor{\bsnm{Nordlund},~\binits{A.}},
  \bauthor{\bsnm{Stein},~\binits{R.F.}}:
\byear{2002}, \batitle{Waves in the magnetized solar atmosphere. i. basic
  processes and internetwork oscillations}. \textit{\bjtitle{Astrophys. J.}}
  \textbf{\bvolume{564}}, \bfpage{508}\,--\,\blpage{524}.
\end{barticle}
\endbibitem

\bibitem[\protect\citeauthoryear{{R{\"{u}}edi} {\it
  et~al.}}{1998}]{Ruedi+Solanki+Stenflo+Tarbell+Scherrer1998}
\begin{barticle}
\bauthor{\bsnm{R{\"{u}}edi},~\binits{I.}},
  \bauthor{\bsnm{Solanki},~\binits{S.K.}},
  \bauthor{\bsnm{Stenflo},~\binits{J.}}, \bauthor{\bsnm{Tarbell},~\binits{T.}},
  \bauthor{\bsnm{Scherrer},~\binits{P.H.}}:
\byear{1998}, \batitle{Oscillations of sunspot magnetic fields}.
  \textit{\bjtitle{Astron. Astrophys.}} \textbf{\bvolume{335}},
  \bfpage{L97}\,--\,\blpage{L100}.
\end{barticle}
\endbibitem

\bibitem[\protect\citeauthoryear{Socas-Navarro, \mbox{Trujillo Bueno}, and
  \mbox{Ruiz Cobo}}{2000}]{Socas-Navarro+etal2000}
\begin{barticle}
\bauthor{\bsnm{Socas-Navarro},~\binits{H.}}, \bauthor{\bsnm{\mbox{Trujillo
  Bueno}},~\binits{J.}}, \bauthor{\bsnm{\mbox{Ruiz Cobo}},~\binits{B.}}:
\byear{2000}, \batitle{Anomalous circular polarization profiles in sunspot
  chromospheres}. \textit{\bjtitle{Astrophys. J.}} \textbf{\bvolume{544}},
  \bfpage{1141}\,--\,\blpage{1154}.
\end{barticle}
\endbibitem

\bibitem[\protect\citeauthoryear{Thomas and
  Stanchfield}{2000}]{Thomas+Stanchfield2000}
\begin{barticle}
\bauthor{\bsnm{Thomas},~\binits{J.H.}},
  \bauthor{\bsnm{Stanchfield},~\binits{D.C.H.}}:
\byear{2000}, \batitle{Fine-scale magnetic effects on p-modes and higher
  frequency acoustic waves in a solar active region}. \textit{\bjtitle{Astrophys. J.}}
  \textbf{\bvolume{537}}, \bfpage{1086}\,--\,\blpage{1093}.
\end{barticle}
\endbibitem

\bibitem[\protect\citeauthoryear{{Tziotziou} {\it
  et~al.}}{2006}]{Tziotziou+etal2006}
\begin{barticle}
\bauthor{\bsnm{Tziotziou},~\binits{K.}},
  \bauthor{\bsnm{Tsiropoula},~\binits{G.}}, \bauthor{\bsnm{Mein},~\binits{N.}},
  \bauthor{\bsnm{Mein},~\binits{P.}}:
\byear{2006}, \batitle{Observational characteristics and association of umbral
  oscillations and running penumbral waves}. \textit{\bjtitle{Astron. Astrophys.}}
  \textbf{\bvolume{456}}, \bfpage{689}\,--\,\blpage{695}.
\end{barticle}
\endbibitem

\bibitem[\protect\citeauthoryear{{Tziotziou} {\it
  et~al.}}{2007}]{Tziotziou+etal2007}
\begin{barticle}
\bauthor{\bsnm{Tziotziou},~\binits{K.}},
  \bauthor{\bsnm{Tsiropoula},~\binits{G.}}, \bauthor{\bsnm{Mein},~\binits{N.}},
  \bauthor{\bsnm{Mein},~\binits{P.}}:
\byear{2007}, \batitle{Dual-line spectral and phase analysis of sunspot
  oscillations}. \textit{\bjtitle{Astron. Astrophys.}} \textbf{\bvolume{463}},
  \bfpage{1153}\,--\,\blpage{1163}.
\end{barticle}
\endbibitem

\bibitem[\protect\citeauthoryear{{Rouppe van~der Voort} {\it
  et~al.}}{2003}]{Rouppe+etal2003}
\begin{barticle}
\bauthor{\bparticle{Rouppe van~der }\bsnm{Voort},~\binits{L.H.M.R.}},
  \bauthor{\bsnm{Rutten},~\binits{R.J.}},
  \bauthor{\bsnm{S{\"u}tterlin},~\binits{P.}},
  \bauthor{\bsnm{Sloover},~\binits{P.J.}},
  \bauthor{\bsnm{Krijger},~\binits{J.M.}}:
\byear{2003}, \batitle{La palma observations of umbral flashes}.
  \textit{\bjtitle{Astron. Astrophys.}} \textbf{\bvolume{403}},
  \bfpage{277}\,--\,\blpage{285}.
\end{barticle}
\endbibitem

\bibitem[\protect\citeauthoryear{{Vecchio} {\it
  et~al.}}{2007}]{Veccio+etal2007}
\begin{barticle}
\bauthor{\bsnm{Vecchio},~\binits{A.}}, \bauthor{\bsnm{Cauzzi},~\binits{G.}},
  \bauthor{\bsnm{Reardon},~\binits{K.P.}},
  \bauthor{\bsnm{Janssen},~\binits{K.}}, \bauthor{\bsnm{Rimmele},~\binits{T.}}:
\byear{2007}, \batitle{Solar atmospheric oscillations and the chromospheric
  magnetic topology}. \textit{\bjtitle{Astron. Astrophys.}} \textbf{\bvolume{461}},
  \bfpage{L1}\,--\,\blpage{L4}.
\end{barticle}
\endbibitem

\bibitem[\protect\citeauthoryear{Vernazza, Avrett, and
  Loeser}{1981}]{Vernazza+Avrett+Loeser1981}
\begin{barticle}
\bauthor{\bsnm{Vernazza},~\binits{J.E.}},
  \bauthor{\bsnm{Avrett},~\binits{E.H.}}, \bauthor{\bsnm{Loeser},~\binits{R.}}:
\byear{1981}, \batitle{Structure of the solar chromosphere. III - models of the
  euv brightness components of the quiet sun}. \textit{\bjtitle{Astrophys. J.}}
  \textbf{\bvolume{45}}, \bfpage{635}\,--\,\blpage{725}.
\end{barticle}
\endbibitem

\bibitem[\protect\citeauthoryear{Zhugzhda}{2007}]{Zhugzhda2007}
\begin{botherref}
\oauthor{\bsnm{Zhugzhda},~\binits{Y.D.}}:
2007, \textit{Astron. Letters} \textbf{in press}.
\end{botherref}
\endbibitem

\bibitem[\protect\citeauthoryear{Zhugzhda and
  Locans}{1981}]{Zhugzhda+Locans1981}
\begin{barticle}
\bauthor{\bsnm{Zhugzhda},~\binits{Y.D.}}, \bauthor{\bsnm{Locans},~\binits{V.}}:
\byear{1981}, \batitle{Resonance oscillations in sunspots}.
  \textit{\bjtitle{Sov. Astron. Lett.}} \textbf{\bvolume{7}}, 25.
\end{barticle}
\endbibitem

\end{thebibliography}
\end{document}